\documentclass[fleqn,usenatbib]{mnras}
\usepackage[T1]{fontenc}
\usepackage{ae,aecompl}

\usepackage{graphicx}	% Including figure files
\usepackage{amsmath}	% Advanced maths commands
\usepackage{amssymb}	% Extra maths symbols
\usepackage{subfig}
\usepackage{newtxtext,newtxmath}
\title[The LOFAR view on hard X-ray GRG]{Hard X-ray selected giant radio galaxies -- III. The LOFAR view}

\author[G. Bruni et al.]{G. Bruni,$^{1}$\thanks{Contact e-mail: \href{mailto:gabriele.bruni@inaf.it}{gabriele.bruni@inaf.it}}
    M. Brienza,$^{2,3}$
    F. Panessa,$^{1}$
    L. Bassani,$^{4}$
    D. Dallacasa,$^{2,3}$
    T. Venturi,$^{3}$
    \newauthor
    R. D. Baldi,$^{3,5}$
    A. Botteon,$^{6}$
    A. Drabent,$^{7}$
    A. Malizia,$^{4}$
    F. Massaro,$^{8,9,10,11}$
    \newauthor
    H. J. A. R\"ottgering,$^{6}$
    P. Ubertini,$^{1}$
    F. Ursini,$^{12}$
    and
    R. J. van Weeren$^{6}$
\\
    $^{1}$INAF - Istituto di Astrofisica e Planetologia Spaziali, via Fosso del Cavaliere 100, I-00133 Roma, Italy \\
    $^{2}$DiFA - Dipartimento di Fisica e Astronomia, Universit\`a di Bologna, via P. Gobetti 93/2, I-40129 Bologna, Italy \\
    $^{3}$INAF - Istituto di Radioastronomia, via Piero Gobetti 101, I-40129 Bologna, Italy \\   
    $^{4}$INAF - Osservatorio di astrofisica e scienza dello spazio di Bologna, Via Piero Gobetti 93/3, 40129 Bologna, Italy \\
    $^{5}$School of Physics and Astronomy, University of Southampton, Southampton, SO17 1BJ, UK\\
    $^{6}$Leiden Observatory, Leiden University, PO Box 9513, 2300 RA Leiden, The Netherlands\\
    $^{7}$Th\"uringer Landessternwarte, Sternwarte 5, D-07778 Tautenburg, Germany\\
    $^{8}$Dipartimento di Fisica, Universit\`a degli Studi di Torino, via Pietro Giuria 1, I-10125 Torino, Italy\\
    $^{9}$INAF - Osservatorio Astrofisico di Torino, via Osservatorio 20, 10025 Pino Torinese, Italy\\ 
    $^{10}$Istituto Nazionale di Fisica Nucleare, Sezione di Torino, I- 10125 Torino, Italy\\ 
    $^{11}$Consorzio Interuniversitario per la Fisica Spaziale, via Pietro Giuria 1, I-10125 Torino, Italy\\   
    $^{12}$Dipartimento di Matematica e Fisica, Universit\`a degli Studi Roma Tre, via della Vasca Navale 84, 00146 Roma, Italy
    }

\date{Accepted XXX. Received YYY; in original form ZZZ}

\pubyear{2015}

\begin{document}
\label{firstpage}
\pagerange{\pageref{firstpage}--\pageref{lastpage}}
\maketitle

%%%%%%%%%%%%%%%%%%%%%%%%%%%%%%%%%%%%%%%%%%%%%%%%%%%%%%%%%%%%%%%%%%%%%%%%%%%%%%%%%%%
%%%%%%%%%%%%%%%%%%%%%%%%%%%%%%%%%%%%%%%%%%%%%%%%%%%%%%%%%%%%%%%%%%%%%%%%%%%%%%%%%%%

\begin{abstract}
Giant radio galaxies (GRGs), with extended structures reaching hundreds of kpc, are among the most spectacular examples of ejection of relativistic plasma from supermassive black holes. In this work, third of a series, we present LOw Frequency ARray (LOFAR) images at 144 MHz, collected in the framework of the LOFAR Two-metre Sky Survey Data Release 2 (LoTSS DR2), for nine sources extracted from our sample of hard X-ray selected GRGs (HXGRG, i.e. from \emph{INTEGRAL}/IBIS and \emph{Swift}/BAT catalogues at >20 keV). 
%%%
Thanks to the resolution and sensitivity of LoTSS, we could probe the complex morphology of these GRGs, unveiling cases with diffuse (Mpc-scale) remnant emission, presence of faint off-axis wings, or a misaligned inner jet. In particular, for one source (B2\,1144+35B), we could clearly detect a $\sim$300 kpc wide off-axis emission, in addition to an inner jet which orientation is not aligned with the lobes axis. For another source (J1153.9+5848) a structure consistent with jet precession was revealed, appearing as an X-shaped morphology with relic lobes having an extension larger than the present ones, and with a different axis orientation. 
%%%
From an environment analysis, we found 2 sources showing an overdensity of cosmological neighbours, and a correspondent association with a galaxy cluster from catalogues.
%%%
Finally, a comparison with radio-selected GRGs from LoTSS DR1 suggested that, on average, HXGRG can grow to larger extents. 
%%%
These results highlight the importance of deep low-frequency observations to probe the evolution of radio galaxies, and ultimately estimate the duty cycle of their jets.     
\end{abstract}


%%%%%%%%%%%%%%%%%%%%%%%%%%%%%%%%%%%%%%%%%%%%%%%%%%
%%%%%%%%%%%%%%%%% BODY OF PAPER %%%%%%%%%%%%%%%%%%
%%%%%%%%%%%%%%%%%%%%%%%%%%%%%%%%%%%%%%%%%%%%%%%%%%

\section{Introduction}

Giant radio galaxies (GRG, \citealt{1974Natur.250..625W}) are among the most extreme manifestations of the accretion/ejection processes on supermassive black holes. During their $\sim$100 Myr timescale evolution \citep{2004AcA....54..249M, 2008MNRAS.385.1286J}, they produce jets of plasma extending hundreds of kpc away from their active galactic nucleus (AGN) core. Their projected linear size, comprised between a conventional threshold of 0.7 Mpc and a maximum of 4.69 Mpc to date \citep{2008ApJ...679..149M}, implies that their jets are able to probe the intergalactic medium, at spatial scales comparable with galaxy clusters. In the past decades, several authors performed detailed studies of the age of GRG, on the basis of their kinematics or plasma radiative losses, dating the ancient lobes as structures formed during tens, if not hundreds, of Myr (see e.g. \citealt{1998A&A...329..431M,2016ApJ...817L...1H,2020MNRAS.495..143C}), eventually becoming remnant radio structures \citep{2015MNRAS.447.2468H,2016A&A...585A..29B}. With such a long activity period, GRG are the ideal testbed to study the duration of the radio phase in AGN, and its duty cycle. Indeed, several examples of GRG with recurrent radio activity have been found by different authors \citep{2004MNRAS.355..845K,2015A&A...584A.112O,2016ApJ...817L...1H,2020MNRAS.494..902B}, either in the form of two symmetric pairs of lobes with respect to the active nucleus (termed double-double radio galaxy, \citealt{2000A&AS..146..293S}), or an X-shaped morphology (misaligned double-double, \citealt{Cheung_2007,2013MNRAS.436..690S}) possibly linked to a change in jet axis, or the presence of a young radio source in the core at odds with the age inferred from the Mpc-scale lobes \citep{2019ApJ...875...88B}.    

With the advent of the new generation low-frequency surveys the number of known GRG has dramatically increased, populating the parameter space of GRG. This suggests that their abundance has been underestimated so far, mainly because of the limited sensitivity to extended emission of the previous generation radio surveys \citep{2021MNRAS.501.3833D}. In particular, LOFAR \citep{2013A&A...556A...2V}, with its unique sensitivity to extended steep-spectrum structures and a typical resolution of 6 arcsec in the MHz domain, is the ideal instrument to study these complex and low surface brightness objects. Indeed, the emission from plasma generated at the beginning of GRG activity shows steep spectra, and therefore its emission is brighter and more easily detected at low frequencies. In the framework of the LOFAR surveys key science project, single objects \citep{2020MNRAS.495..143C,2019A&A...628A..69S}, as well as samples of GRG \citep{2020A&A...635A...5D,2020A&A...638A..48S} have been studied with unprecedented detail, allowing to extend the census to hundreds of sources. As a whole, morphological studies at low frequencies, through synthesis imaging with radio interferometers, can help understanding the plasma dynamics. This information can then be used to investigate the duty cycle of the radio phase in AGN.   

In the past few years, the GRACE\footnote{\href{https://sites.google.com/inaf.it/grace/}{https://sites.google.com/inaf.it/grace/}} project (Giant RAdio galaxies and their duty CyclE) has targeted GRG selected by cross-correlating hard X-ray catalogues ($>$20 keV) from \emph{Swift}/BAT and \emph{INTEGRAL}/IBIS with radio catalogues (the NRAO Very Large Array Sky Survey - NVSS, \citealt{1998AJ....115.1693C} - and the Sydney University Molonglo Sky Survey - SUMSS, \citealt{1999AJ....117.1578B}). As a first result of this selection, \cite{2016MNRAS.461.3165B} found that the fraction of GRG in the parent sample of 64 radio galaxies was about four times larger than in previous radio-only studies ($\sim$25\% vs the classical $\sim$6\% from the 3CR catalogue, see \citealt{1999MNRAS.309..100I} and 8\% in the 3CRR sample, see \citealt{1983MNRAS.204..151L}). We thus embarked on a multi-wavelength study of the 15 GRG from the \cite{2016MNRAS.461.3165B} sample (see \cite{2020MNRAS.494..902B} for the full sample description), aimed at understanding the reason for such a correlation. In the first paper of this collection, \cite{2018MNRAS.481.4250U} showed how the radio luminosity of the lobes and jet power are dimmer than what expected from the X-ray power of the core, suggesting a recently restarted activity. This was later on confirmed by \cite{2019ApJ...875...88B,2020MNRAS.494..902B}, either on the basis of the radio morphology, or for the presence of a recently restarted radio source in the core. 

In this paper, for the first time, we present LOFAR images of 9 HXGRG, aiming at characterising their large scale morphology and signs of restarted radio activity. Moreover, we perform a comparison between HXGRG and a sample of radio-selected GRGs, to investigate systematic differences in their main properties. Recently, \cite{2021MNRAS.500.3111B} presented an update of the sample, adding 8 GRG plus a candidate source. We present here five GRG from the original sample, plus another four from the catalogue update.
In the following, we adopt the convention $S\propto\nu^{\,\alpha}$ for the spectral index $\alpha$, and the same cosmological parameters as in \cite{2016MNRAS.461.3165B} for consistency, i.e. $H_{0}$=$71$ km/s/Mpc, $\Omega_{\rm{m}}$=$0.27$, and $\Omega_{\Lambda}$=$0.73$.

%%%%%%%%%%%%%%%%%%%%%%%%%%%%%%%%%%%%%%%%%%%%%%%%%%%%%%%%%%%%%%%%%%%

\begin{table}
    \centering
    \begin{tabular}{lccccccc}
    \hline
    Source              &  RMS (6$''$)         &  RMS (20$''$)        \\
                        &  ($\mu$Jy/beam)   & ($\mu$Jy/beam)    \\
    \hline             
    J0318+684           &  220              & 360           \\
    J0801.7+4764        &  130              & 420           \\    
    B2\,1144+35B        &   90              & 180           \\
    J1153.9+5848        &  100              & 160           \\
    J1238.4+5349        &   90              & 240           \\
    J1503.7+6850        &  150              & 390           \\
    4C+63.22            &  120              & 360           \\
    Mrk\,1498           &  110              & 220           \\
    4C+34.47            &  230              & 760           \\   
    \hline
    \end{tabular}
    \caption{Summary of LoTSS data presented in this work, with typical RMS noise obtained for the images convolved at 6$''$ and 20$''$ resolution.}
    \label{tab:LOTSS}
\end{table}

%%%%%%%%%%%%%%%%%%%%%%%%%%%%%%%%%%%%%%%%%%%%%%%%%%%%%%%%%%%%%%%%%%%

\begin{table}
    \centering
    \begin{tabular}{cccccccc}
    \hline
    Source          &   Code            &  Frequency    &  Beam                         &  RMS              \\
                    &                   &  (GHz)        &  (arcsec)                     &  ($\mu$Jy/beam)   \\
    \hline                                               
    J0318+684       &   20A-476         & 3             &  20                           & 200               \\
    Mrk 1498        &   15B-145         & 6             &  35$\times$15                 & 60                \\
    \hline
    \end{tabular}
    \caption{Summary of VLA data presented in this work, from our observations (20A-476) and archive (15B-145).}
    \label{tab:VLA}
\end{table}

%%%%%%%%%%%%%%%%%%%%%%%%%%%%%%%%%%%%%%%%%%%%%%%%%%%%%%%%%%%%%%%%%%%%

\begin{table*}
    \centering
    \begin{tabular}{lccccccc}
    \hline
    Source             & RA             & DEC          & $z$           & Spatial scale     & LS         &  Total flux density   &  Total radio power \\
                       & (J2000)        & (J2000)      &               & (kpc/arcsec)      & (kpc)      & (Jy)                  &  Log(W/Hz)         \\
    \hline                                                                                                                           
    J0318+684          & 03:18:19.1	    &   +68:29:32  &   0.09            & 1.659          & 1620      & 5.67$\pm$1.13         & 26.03             \\
    J0801.7+4764       & 08:01:31.9     &   +47:36:15  &   0.16            & 2.729          & 1260      & 0.46$\pm$0.09         & 25.47             \\
    B2\,1144+35B       & 11:47:22.3     &   +35:01:09  &   0.06            & 1.144          &  860      & 1.46$\pm$0.29         & 25.08             \\
    J1153.9+5848       & 11:53:23.9     &   +58:31:41  &   0.20            & 3.268          &  900      & 0.26$\pm$0.05         & 25.43             \\
    J1238.4+5349       & 12:38:07.7     &   +53:25:56  &   0.35            & 4.907          &  950      & 0.52$\pm$0.10         & 26.26             \\
    J1503.7+6850       & 15:04:13.1     &   +68:56:12  &   0.32            & 4.622          & 1230      & 2.29$\pm$0.46         & 26.82             \\
    4C+63.22           & 15:23:45.8	    &   +63:39:24  &   0.20            & 3.268          &  790      & 3.30$\pm$0.66         & 26.53             \\
    Mrk\,1498          & 16:28:03.8	    &   +51:46:30  &   0.05            & 0.965          & 1150      & 2.60$\pm$0.52         & 25.16             \\
    4C+34.47           & 17:23:20.6	    &   +34:18:00  &   0.21            & 3.395          & 1120      & 5.21$\pm$1.04         & 26.77             \\ 
    \hline
    \end{tabular}
    \caption{Quantities for the sample of 9 HXGRG presented in this work, including projected linear sizes, total flux densities, and total radio power, as measured from the LOFAR images at 144 MHz (20$''$ resolution).}
    \label{tab:LOFAR}
\end{table*}

%%%%%%%%%%%%%%%%%%%%%%%%%%%%%%%%%%%%%%%%%%%%%%%%%%%%%%%%%%%%%%%%%%%%%%%%%%%%%%%%%%%%

\section{Radio data}

\subsection{LOFAR}
We made use of data from the LOFAR Two-metre Sky Survey (LoTSS, \citealt{2017A&A...598A.104S}) Data Release 1 and the upcoming 2 (DR1 and DR2 hereafter), carried out at 144 MHz (frequency range 120-168 MHz). LoTSS DR1 \citep{2019A&A...622A...1S} covered an area of 424 square degrees in the Hobby-Eberly Telescope Dark Energy Experiment (HETDEX) spring field, between right ascension 10h45m00s to 15h30m00s and declination 45$^\circ$00$'$00$''$ to 57$^\circ$00$'$00$''$, reaching a median sensitivity of 71 $\mu$Jy/beam at 6 arcsec angular resolution. LoTSS DR2 (Shimwell et al. in prep.) will offer additional coverage on two fields centered at a Galactic latitude of 0h and 13h, for a total of 5700 square degrees, and processed with an improved version of the pipeline.
Among the nine sources analysed in this work, only one (J1238.4+5349) is located in the HETDEX region, and therefore was already published as part of the LoTSS DR1, while the remaining eight are located in the sky region covered by the DR2, and are presented here for the first time. All datasets used in this work have been processed by the Survey KSP team using the standard DR2 approach (Shimwell et al. in prep.). This includes the use of an upgraded version of the LoTSS direction dependent pipeline (\citealt{2018A&A...611A..87T,2020arXiv201108328T}), which leads to significant improvements of the final images with respect to the DR1. 

To further refine the imaging in the direction of the nine targets, we proceeded as follows. First, we subtracted out of the visibilities all the sources outside the region containing the target, and performed additional phase and amplitude self-calibration loops in the target direction, following the same procedure as previous works from the LoTSS collaboration \citep{2020arXiv201102387V}. We then used these datasets to produce the final images running the {\tt{WSCLEAN}} imager \citep{offringa-wsclean-2014} on the Bologna node of the LOFAR-IT infrastructure, fine-tuning different sets of parameters with the goal of enhancing the reconstruction of the diffuse and faint emission of the sources. Final images were obtained with two circular Gaussian restoring beams of 6 and 20 arcsec for each source, to exploit both the full resolution of the array and a tailored large beam to better detect the diffuse emission. For both resolutions we used a Briggs weighting scheme with robust=--0.5 and multi-scale cleaning (see \citealt{2017MNRAS.471..301O}. Additionally, we applied a Gaussian taper to obtain 20-arcsec angular resolution images.

Table \ref{tab:LOTSS} summarizes the final typical RMS noise obtained for the different images, at both resolutions, and for each source. Analogously to \cite{2019A&A...622A...1S}, and to recent works on GRGs performed with LoTSS data \citep{2020A&A...635A...5D}, we assumed a conservative error of 20\% for the measured flux densities.

%%%%%%%%%%%%%%%%%%%%%%%%%%%%%%%%%%%%%%%%%%%%%%%%%%%%%%%%%%%%%%%%%%%

\subsection{VLA}

For two sources (J0318+684 and Mrk\,1498), we could consider additional observations from the Karl G. Jansky Very Large Array (VLA, see Tab. \ref{tab:VLA}). For J0318+684, DDT observations were obtained, and performed on May 24, 2020 (project 20A-476) with the array in C configuration at 3 GHz. The on-source time was 15 min. For Mrk\,1498, we made use of archival data (project 15B-145). Observations at 6 GHz were performed on November 11, 2015, in D configuration, for a total on-source time of 36 min.

Data were reduced with the {\tt{CASA}}\footnote{\href{https://casa.nrao.edu/}{https://casa.nrao.edu/}} software \citep{2007ASPC..376..127M}, running VLA pipeline for initial calibration. Then, target data were split and imaging performed with the {\tt{TCLEAN}} task, through two iterations of phase-only self-calibration. For J0318+684, almost 40\% of the bandwidth was flagged out for the 3 GHz observations, due to radio frequency interference. A final RMS noise of 200 $\mu$Jy/beam was obtained. For Mrk\,1498, in order to mitigate the lack of sensitivity of the VLA to structures larger than 4 arcmin (in D configuration and at 6 GHz) we made use of the lower 25\% of the receiver band only (4-5 GHz), recovering as much as possible the extended emission of the target. This resulted in a final RMS noise of 60 $\mu$Jy/beam.

%%%%%%%%%%%%%%%%%%%%%%%%%%%%%%%%%%%%%%%%%%%%%%%%%%%%%%%%%%%%%%%%%%%%%%%%%%
%%%%%%%%%%%%%%%%%%%%%%%%%%%%%%%%%%%%%%%%%%%%%%%%%%%%%%%%%%%%%%%%%%%%%%%%%%

\section{Results}
\label{sec:results}

In the following, we present the morphology revealed by LOFAR images for 9 HXGRG extracted from \cite{2016MNRAS.461.3165B} or \cite{2021MNRAS.500.3111B}. In the figures, the 6 arcsec resolution image is plotted in color scale, while the contours are from the 20 arcsec resolution one, unless specified differently. For a few sources, we also provide 
a zoom of the core region. The X-ray position of the core from \cite{2020MNRAS.494..902B} or \cite{2021MNRAS.500.3111B} is indicated with a cross. In table \ref{tab:LOFAR} we report quantities extracted from LOFAR images, including updated values for the projected linear sizes (estimated from the angular size between the outermost 3$\times$RMS contours), and total flux densities from the LOFAR 20 arcsec resolution images. 

%%%%%%%%%%%%%%%%%%%%%%%%%%%%%%%%%%%%%%%%%%%%%%%%%%%%%%%%%%%%%%%%%%%

\begin{figure*}
\centering
  \includegraphics[width=\textwidth]{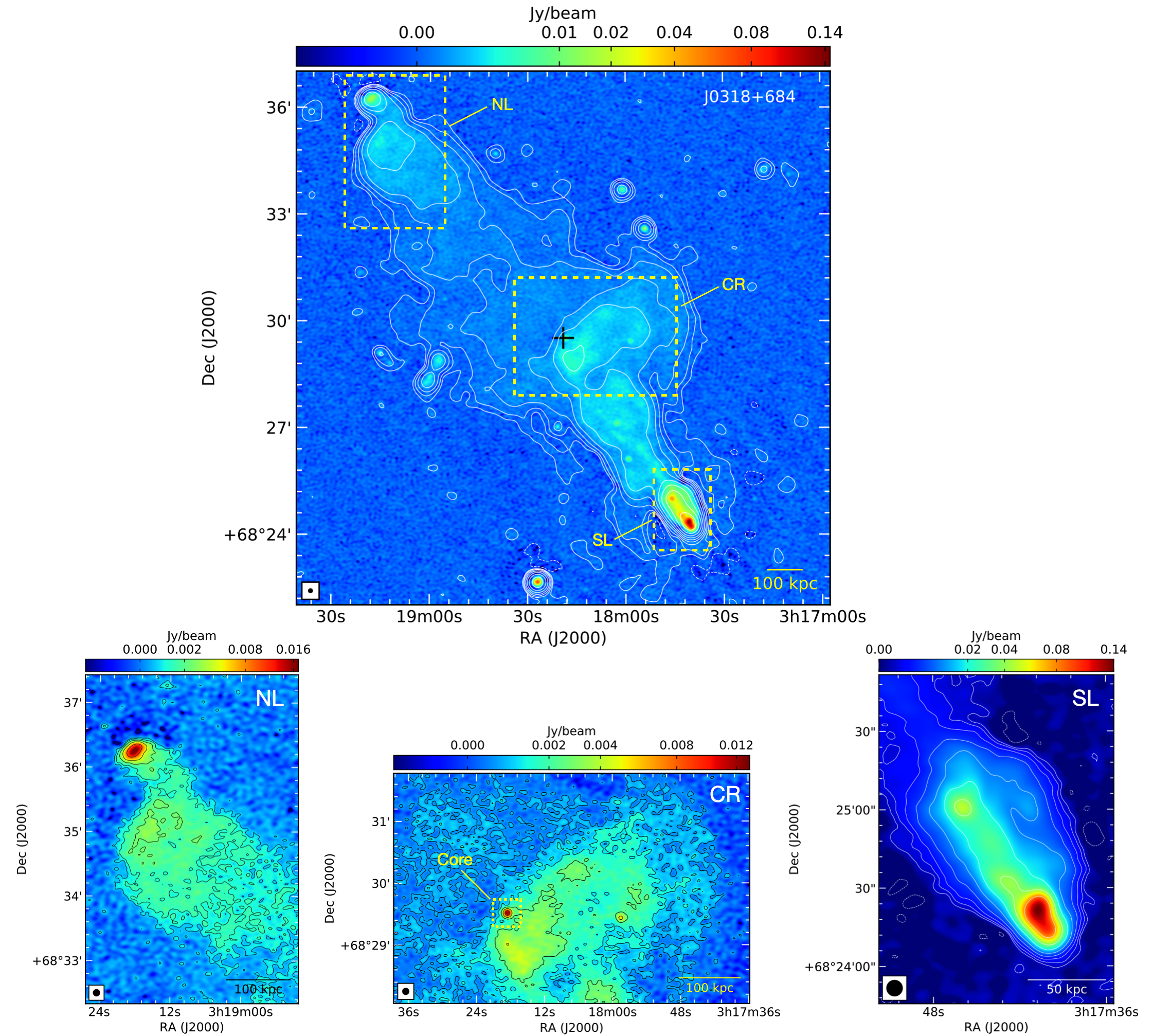}
\caption{Top panel: LOFAR image of the source J0318+684. The color scale represents the 6$''$ resolution image, while contours are from the 20$"$ one. Bottom panels: zoom on the Northern lobe (left), core region (middle), and Southern lobe (right) from the LOFAR image at 6$''$ resolution. Contours are multiples of the image RMS, namely 3$\times$RMS$\times$(--1, 1, 2, 4, 8, 16, 32, 64, 128, 256). Negative contours are dashed. The beam is shown in the lower-left corner.}
\label{fig:0318_LOFAR}
\end{figure*}

%%%%%%%%%%%%%%%%%%%%%%%%%%%%%%%%%%

\begin{figure*}
\centering
  \includegraphics[width=\textwidth]{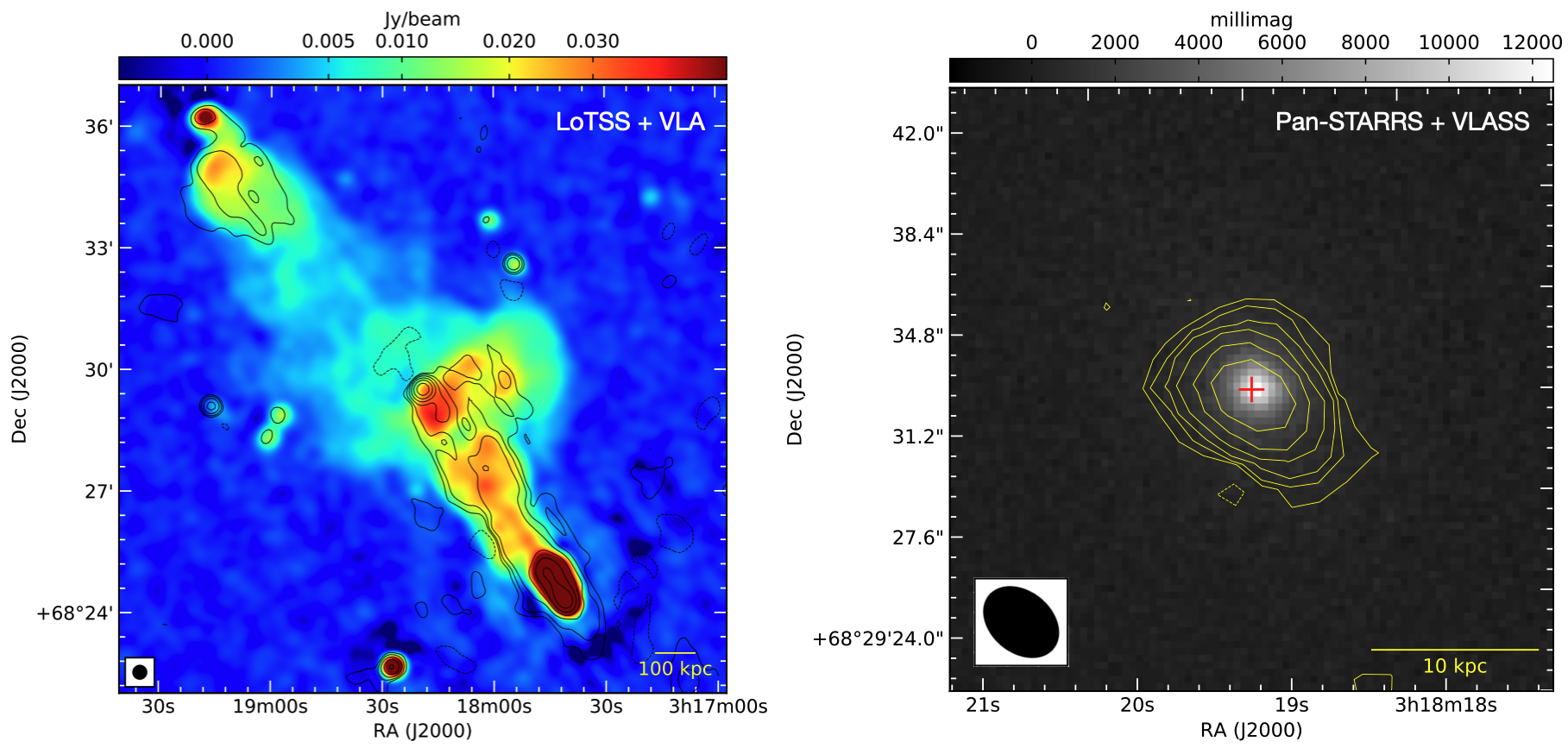}
\caption{J0318+684. Left panel: comparison between LOFAR (144 MHz, in colors) and VLA (3 GHz, in contours) images at a matched resolution of 20$''$. The beam is shown in the lower-left corner. Right panel: Pan-STARRS i-band image of the host galaxy, with VLASS countours superimposed (2--4 GHz, $3''\times2''$ angular resolution). Contours are multiples of the image RMS, namely 3$\times$RMS$\times$(--1, 1, 2, 4, 8, 16, 32, 64, 128, 256), negative contours are dashed. The VLASS beam is shown in the lower-left corner.}
\label{fig:0318_PAN+VLASS+VLA}
\end{figure*}

%%%%%%%%%%%%%%%%%%%%%%%%%%%%%%%%%%%%%%%%%%%%%%%%%%%%%%%%%%%%%%%%%%%

\subsection*{J0318+684}

This is one of the most extended GRG of our sample (1.5 Mpc, \citealt{2016MNRAS.461.3165B}). In our previous works, we discussed the discovery of a young radio source (GigaHertz Peaked Spectrum source - GPS) in the core region, suggesting a recent reactivation of the jet \citep{2019ApJ...875...88B}, and discussed the X-shaped like morphology \citep{2020MNRAS.494..902B}. Beyond our proprietary data, in the discussion of this object we considered other data available in the literature. Namely, Very Large Array Sky Survey data at 3 GHz (VLASS, first epoch quick-look image, \citealt{2020PASP..132c5001L}), and optical images from The Panoramic Survey Telescope and Rapid Response System (Pan-STARRS, \citealt{chambers2016panstarrs1}). 

In Fig. \ref{fig:0318_LOFAR} (top panel) we show the LOFAR full resolution image of the source, with the low resolution image contours overlaid. The source morphology is consistent with our GMRT image at 325 MHz from \cite{2020MNRAS.494..902B}. The Northern and Southern lobes are the brightest components, with the latter having a peak surface brightness of 143 mJy/beam with respect to 17.3 mJy/beam for the Northern one (at 6$''$ resolution). The core is well visible as a compact and bright spot at the center of the source, with a peak of 13.1 mJy/beam. Diffuse, off-axis emission is present, with a remarkable extension of $\sim$3 arcmin towards NW, and $\sim$2.5 arcmin SE from the core (corresponding to 300 kpc and 250 kpc, respectively). Overall, more diffuse emission between core and lobes is visible with respect to our previous GMRT observations. Zoomed images of the lobes and core regions are reported in Fig. \ref{fig:0318_LOFAR} (bottom panels). Both the Northern and Southern lobes present a hotspot, typical of the FRII morphology, with an enhanced surface brightness due to the interaction of the jet plasma with the intergalactic medium (IGM). This is a clear sign of a recent replenishment of plasma, indicating that the jets are still active or, from kinematic considerations, that the radio phase fueling the lobes has stopped a maximum of $T<LS/2/v$ kyr ago, where $v$ is the average jet velocity (typically $\sim$0.1$c$), and $LS$ is the linear size of the GRG in kpc (1500 kpc). This roughly correspond to a $T<24$ Myr. The Southern lobe shows a knotty structure in the region closest to the edge ($\sim$100 kpc, see Fig. \ref{fig:0318_LOFAR}, bottom-right panel). The off-axis extended emission, already highlighted by our GMRT map at 325 MHz from \cite{2020MNRAS.494..902B}, is even more evident in the LOFAR image, as expected from the very steep ($<$-1) spectral index discussed in our previous work. 

In Fig. \ref{fig:0318_PAN+VLASS+VLA} (left panel) we show the comparison between LOFAR (in colors) and our VLA image (3 GHz, contours), restored with a matched angular resolution of 20 arcsec: the off-axis emission is only partially detected by the VLA (due to its lower sensitivity to extended structures, and the steep-spectral index), while the Southern jet/lobe and the core are evident. Analogously to the LOFAR image, the Northern lobe is fainter with respect to the Southern one. The first half of the structure connecting the core to the Northern lobe is not detected. Finally, observations at 6 GHz and 10 GHz were designed to image the core region, that was detected as a compact component. However, the limited resolution and UV coverage of these short DDT observations did not allow us to resolve the core at the kpc scale, we thus make use of VLASS data for this purpose (see below). 
In Fig. \ref{fig:0318_PAN+VLASS+VLA} (right panel) we present a zoom into the core region, making use of the Pan-STARRS i-band image (in gray), tracing the host galaxy structure, and VLASS data (contours), showing the radio emission with better detail than our VLA observations. Indeed, at an angular resolution of 3$\times$2 arcsec, VLASS contours allow us to see an elongation towards the south-west direction, suggesting an inner jet orientation compatible with the Mpc-scale jet one. The new radio phase, previously identified through a GPS, has thus most probably the same jet axis as the older one. At the redshift of the source, the deconvolved size of the radio core as measured from the VLASS image is 0.9$\times$0.4 arcsec, at a position angle of $\sim$50 degrees, corresponding to a projected linear size of 1.5$\times$0.7 kpc. This size is compatible with the one expected for GPS sources ($\le$1 kpc, \citealt{1998PASP..110..493O}). This is also confirmed by archival Very Long Baseline Array (VLBA) observations at pc-scale resolution (Bruni et al. prep.).

In a recent work presenting MeerKAT observations, \cite{2020MNRAS.495.1271C} has shown hydrodynamical backflow in action in a GRG in common with our sample (PKS\,2014--55, see \citealt{2020MNRAS.494..902B}), underlying how the host galaxy position angle (PA), and thus its interstellar medium, can originate the deflection when they are tilted with respect to the jet axis. From a simple bidimensional Gaussian fit of the optical emission from the Pan-STARRS image, we estimated a position angle of 31$\pm$7 degrees for the host galaxy major axis, consistent with the one measured for the Southern jet axis ($\sim$34 degrees): this alignment between jet and host galaxy major axis does not support the mentioned backflow scenario. Alternatively, \cite{2016A&A...587A..25G} showed how the secondary radio structure of X-shaped radio galaxies is often oriented towards the minor axis of the host galaxy, proposing that it originates from the expansion of the radio cocoon in an asymmetric external medium. This scenario is compatible with the previously discussed orientation of the radio and optical counterparts of this source (based on the jet axis and host galaxy major axis orientation, respectively), suggesting that the diffuse emission perpendicular to the jet axis could have this origin.   

%%%%%%%%%%%%%%%%%%%%%%%%%%%%%%%%%%%%%%%%%%%%%%%%%%%%%%%%%%%%%%%%%%%%

\begin{figure}
    \includegraphics[width=8.5cm]{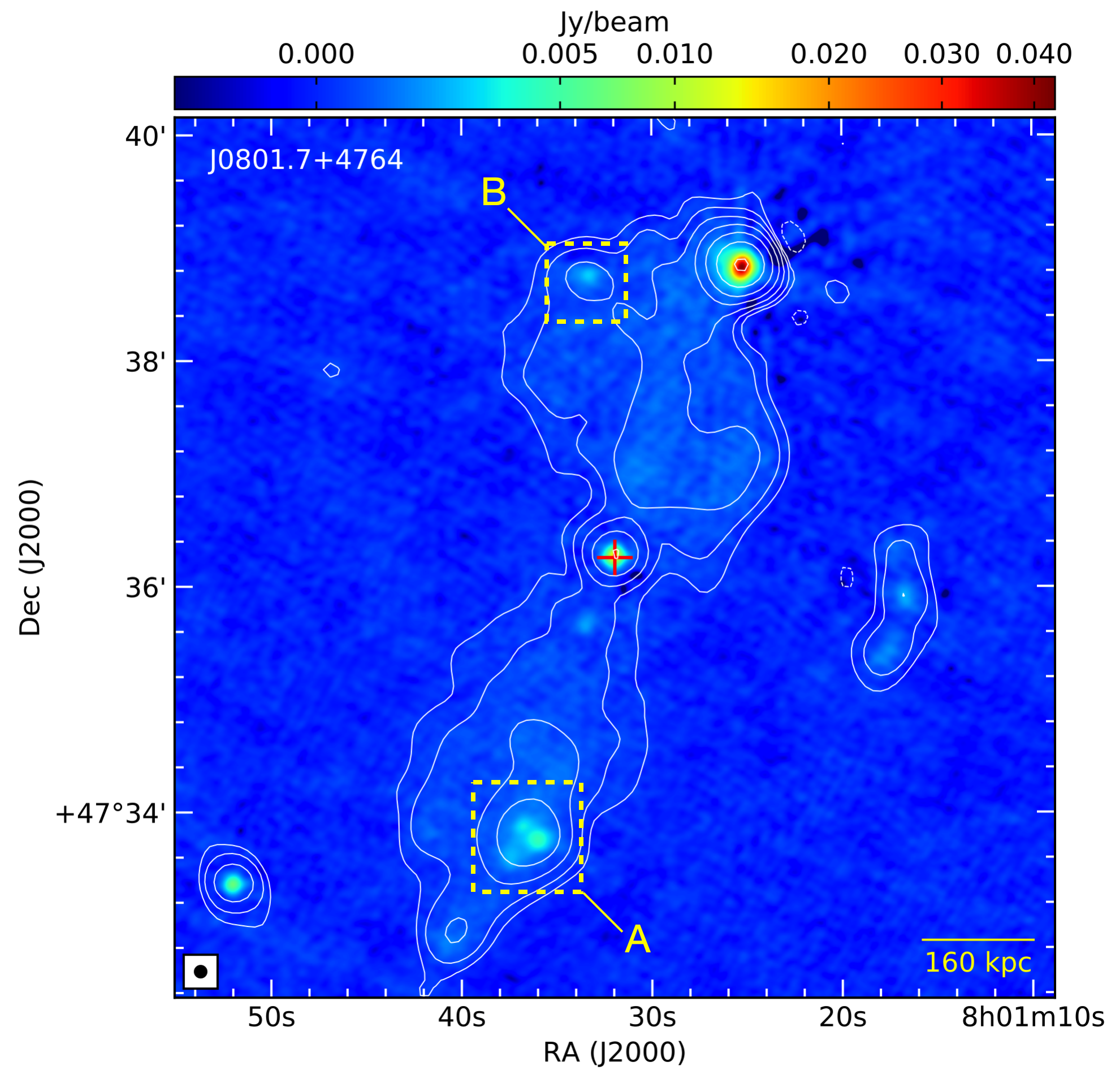}
    \caption{LOFAR image of the source J0801.7+4764. The color scale represents the 6$''$ resolution image, while contours are from the 20$''$ one. Contours are multiples of the image RMS, namely 3$\times$RMS$\times$(--1, 1, 2, 4, 8, 16, 32, 64, 128, 256). Negative contours are dashed. The cross indicates the X-ray position. The 6$''$ beam is shown in the bottom-left corner.}
    \label{fig:0801}
\end{figure}

%%%%%%%%%%%%%%%%%%%%%%%%%%%%%%%%%%%%%%%%%%%%%%%%%%%%%%%%%%%%%%%%%%%%

\subsection*{J0801.7+4764}

The LOFAR image presented here (Fig. \ref{fig:0801}) shows a typical FRII morphology, with the Northern lobe brighter than the Southern one. The Southern lobe presents a brighter region composed by three knots at $\sim$2.5 arcmin from the core (marked as A in the figure). A check with Pan-STARRS images does not show correspondence with any host galaxy, suggesting that these are either real features along the jet, or radio emission from distant galaxies which faint emission is not detected in the optical band (see e.g. \citealt{baldi13}). The same is true for the bright spot westwards of the Northern hotspot (marked as B in the figure), that lies on a region of diffuse emission between the core and the jet. Both A and B component have a spectral index $\alpha_{0.14}^{1.44}=-0.91\pm0.06$, calculated between LOFAR and NVSS data, as expected for diffuse regions in the lobes.
\cite{2012MNRAS.426..851K} provided an estimate of the jet inclination angle with respect to the line of sight, on the basis of the lobes flux density ratio, and assuming it is due to Doppler boosting: they obtained a value of 37$^\circ$, larger than the typical maximum angle of $\sim$10$^\circ$ found for blazars, but still resulting in an asymmetry in jets brightness. Following their argument, the Northern lobe is the one oriented towards the observer, while the Southern lobe is the receding one. 
Information about jet axis inclination allows us to provide an improved estimate of the physical size of the source. The source size as measured from LOFAR observations, from lobe to lobe, is 440 arcsec, that translates into a projected linear size of 1.2 Mpc at the redshift of the source. Considering an inclination angle of 37$^\circ$ for the jet axis, the deprojected linear size of the source is 2 Mpc. In appendix \ref{Appendix:1} we report the serendipitous discovery of a radio galaxy in the LoTSS field of this GRG.   

%%%%%%%%%%%%%%%%%%%%%%%%%%%%%%%%%%%%%%%%%%%%%%%%%%%%%%%%%%%%%%%%%%%

\begin{figure*}
\centering
  \includegraphics[width=\textwidth]{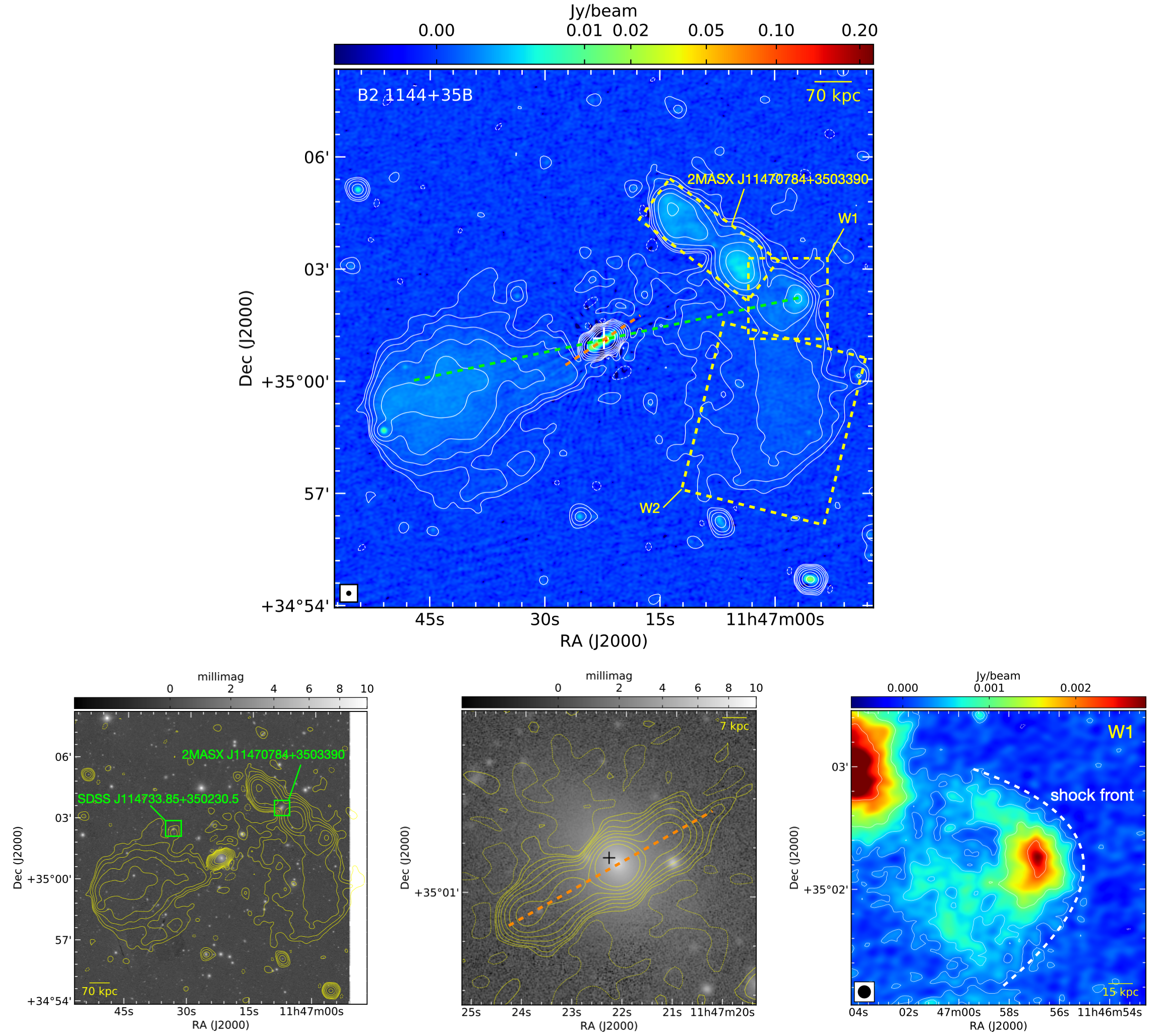}
\caption{Top panel: LOFAR images of source B2\,1144+35B. The color scale represents the 6$''$ resolution image, while contours are from the 20$''$ one. The jet axis connecting the core to component W2 is indicated as a red dashed line, while the axis of the inner jet as an orange dashed line. The LOFAR 6$''$ beam is shown on the lower-left corner.  The cross indicates the X-ray position. Bottom-left panel: overlay of the LOFAR 20$''$ contours with the Pan-STARRS i-band image. The position of the nearby galaxy 2MASX\,J11470784+3503390 is indicated as a red box. Bottom-middle and bottom-right panels: zoom on the core region (LOFAR 6$''$ contours overlaid to Pan-STARRS i-band image) and on the W1 component (both colors and contours from LOFAR 6$''$ image), respectively. The LOFAR 6$''$ beam is shown on the lower-left corner. For all images contours are multiples of the image RMS, namely 3$\times$RMS$\times$(--1, 1, 2, 4, 8, 16, 32, 64, 128). Negative contours are dashed.}
\label{fig:1144}
\end{figure*}

%%%%%%%%%%%%%%%%%%%%%%%%%%%%%%%%%%%%%%%%%%%%%%%%%%%%%%%%%%%%%%%%%%%%

\begin{figure*}
    \includegraphics[width=\textwidth]{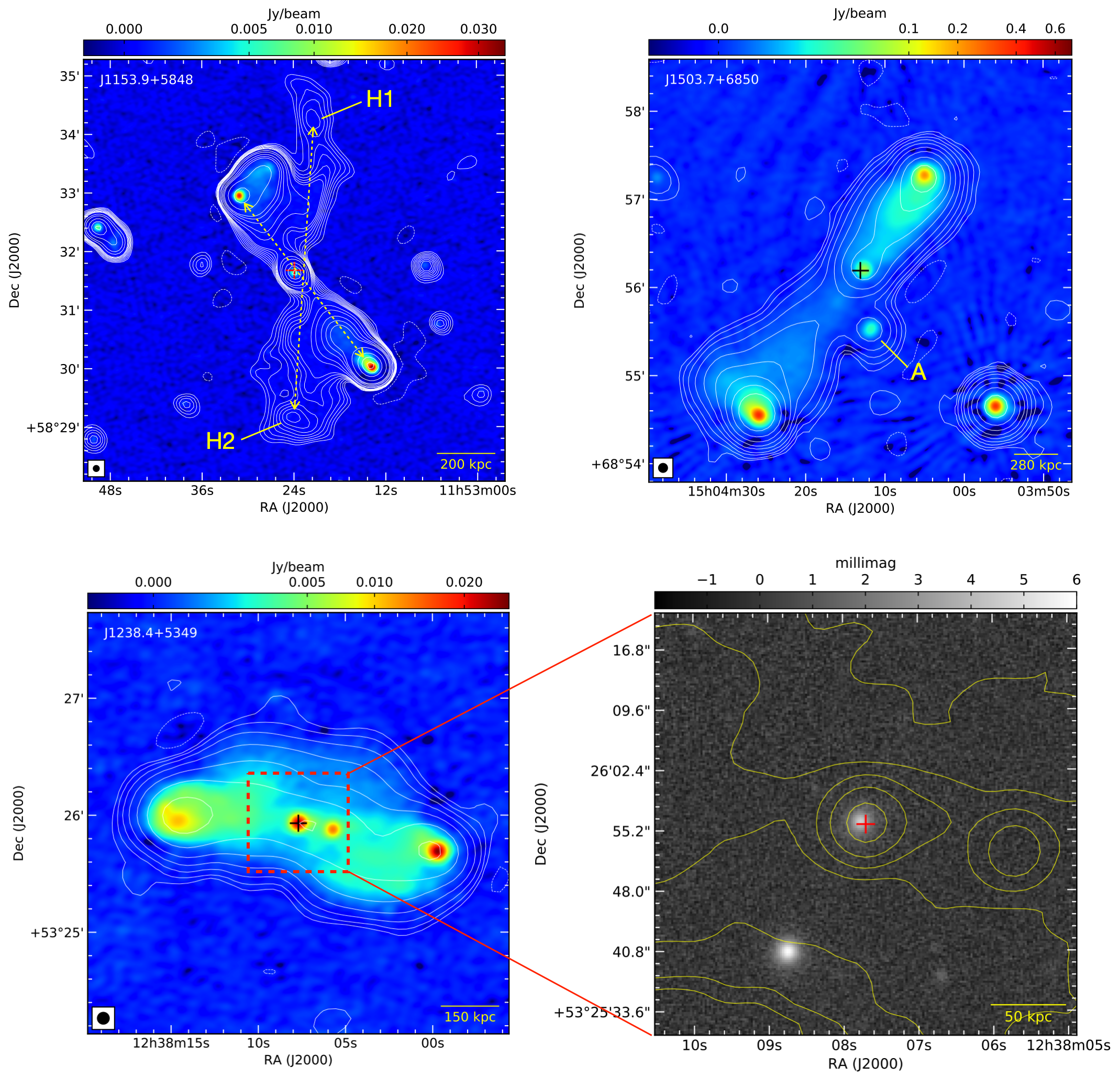}
    \caption{LOFAR images of sources J1153.9+5848, J1503.7+6850, and J1238.4+5349. The color scale represents the 6$''$ resolution image, while contours are from the 20$''$ one.
    Contours for J1153.9+5848 are the image RMS$\times$(--3, 3, 4, 5, 6, 7, 8, 9, 10, 15, 20, 30, 40, 50, 100, 200, 300, 400, 500), while for sources J1503.7+6850 and J1238.4+5349 are 3$\times$RMS$\times$(--1, 1, 2, 4, 8, 16, 32, 64, 128, 256). Negative contours are dashed. The 6$''$ beam is shown in the lower-left corner. Bottom-right panel: zoom on the core region of J1238.4+5349, with LOFAR contours at 6$''$ resolution overlaid to the Pan-STARRS i-band image. The cross indicates the X-ray position.}
    \label{fig:4maps}
\end{figure*}

%%%%%%%%%%%%%%%%%%%%%%%%%%%%%%%%%%%%%%%%%%%%%%%%%%%%%%%%%%%%%%%%%%%

\subsection*{B2\,1144+35B}
\cite{1999ApJ...522..101G,2007A&A...474..409G} performed a detailed study of this source through multi-scale observations, from pc to kpc. In particular, they found discontinuities in the jet morphology through VLA observations, suggesting an intermittent activity. The LOFAR image at 20$''$ resolution (Fig. \ref{fig:1144}, top panel) highlights diffuse, extended emission around the lobes, and, in particular, southwards of the Western one (indicated as W2 in figure \ref{fig:1144}). The latter was partially reconstructed by the VLA image at 1.4 GHz in \cite{1999ApJ...522..101G}, and tentatively identified as the Western lobe. The LOFAR image presented here recovers a larger emission region, both in NS and EW directions, at a frequency a factor of ten lower. We could estimate the spectral index of the W2 region using LOFAR and NVSS data, although in the latter only the brightest part is detected (see \citealt{2020MNRAS.494..902B}): we obtained a value of $\alpha_{0.14}^{1.44}=-0.92\pm0.06$. This is consistent with the value estimated by \cite{1999ApJ...522..101G} between 1.4 and 5 GHz, of about --1. In the NW sector a radio galaxy is present (corresponding to object 2MASX\,J11470784+3503390), moreover at the same redshift as our target ($z=0.06$), and partially overlapping with its line of sight. Indeed, the host galaxy of our target matches with one of the clusters from the \cite{2012MNRAS.422...25S} catalogue (identified with ID 2259 in that work, see Sec. \ref{cluster}). A further galaxy, named SDSS\,J114733.85+350230.5, also at the same redshift as our GRG, is located North-East of the core and is partially overlapping with the Eastern lobe (see Fig. \ref{fig:1144}, bottom-left panel). 

Despite the complexity of the field, the Western lobe of our GRG is well recognizable thanks to the bow-shock visible around the hotspot (indicated as W1 in figure \ref{fig:1144}, top panel), meaning that the jet is expanding in a dense environment, and allowing to draw the jet axis back to the core. A zoom on feature W1 is presented in Fig. \ref{fig:1144} (bottom-right panel). Similarly to what done for W2, we estimated a spectral index of $\alpha_{0.14}^{1.44}=-0.73\pm0.06$ for this component, consistent with the value of about --0.8 estimated by \cite{1999ApJ...522..101G} between 1.4 and 5 GHz. The inner region of the source ($\sim$50 arcsec, corresponding to a projected linear size of $\sim$60 kpc) is resolved in the 6 arcsec resolution image (see Fig. \ref{fig:1144}, bottom-center panel), showing a structure with multiple components: a core (the brightest one), coincident with the optical position of the host galaxy, as well as with the X-ray counterpart of the AGN \citep{2020MNRAS.494..902B}, and three knots along the jets. The axis connecting these components has a position angle of $\sim-60^{\circ}$, while the one traced between the core and the Western lobe (W1) has a position angle of $\sim-77^{\circ}$ (see top panel of Fig. \ref{fig:1144}). With a misalignment between the two of $\sim$17 degrees, a counter-clock reorientation of the jet axis on a Myr timescale can be supposed. Interestingly, an S-shaped morphology of the Eastern jet was shown by the VLA observations at 4.9 GHz presented in \cite{1999ApJ...522..101G}: this is typically expected from jet precession (see \citealt{2019MNRAS.482..240K,2020MNRAS.499.5765H} and references therein) and would support the reorientation discussed above. Nevertheless, none of these two axis seem to be aligned with the hotspot visible in the Eastern lobe. 

We can compare this source with a similar case of GRG reported by \cite{2006ApJ...636..172S}, and termed fat-double due to the noticeable transverse extension of the lobes with respect to the lobe-core distance (SGRS\,J0515--8100). Those authors suggested that the presence of distorted lobes could originate from an intermittent jet activity, with multiple axis changes, due to gravitational perturbations from a companion galaxy. Indeed, the optical counterpart of our target presents a boxy shape, as pointed out by previous authors \citep{1999ApJ...522..101G}, possible aftermath of a merger event. Moreover, the correspondence with a galaxy cluster mentioned above suggests that interactions within the cluster members could be ongoing, making the jet evolution more complex than the case of an isolated radio galaxy. Another example of source with a morphology comparable with this GRG was presented by \cite{2019MNRAS.484.3376L}. Those authors, through an analysis of X-ray cavities and radio jet axis in 3C\,88, concluded that a jet reorientation during the last $\sim10^7$ years is the most likely explanation for the observed radio structure. As a whole, the LOFAR images presented here are consistent with a fat-double morphology, suggesting either multiple radio phases with different jet axes, or a jet reorientation on a Myr timescale.    

%%%%%%%%%%%%%%%%%%%%%%%%%%%%%%%%%%%%%%%%%%%%%%%%%%%%%%%%%%%%%%%%%%%

\subsection*{J1153.9+5848}
 From the LOFAR 20$''$ resolution contours of this source (Fig. \ref{fig:4maps}, top left panel) two symmetric tails of faint emission are detected for the first time, towards both the North and South sides with respect to the core, and with an extension larger than the one of the ongoing radio phase. The contours highlight the presence of bright spots at the end of the tails (named H1 and H2 in Fig. \ref{fig:4maps}, top left panel). The axis connecting the peaks at the edges of the structure intercepts the core region, and has a PA of $\sim$--4$^\circ$. As a comparison, the axis connecting the main lobes of the source shows a PA of $\sim$+36$^\circ$. Both axes are shown as dashed lines in Fig. \ref{fig:4maps}. The fact that the distance between these two features is larger than the one between the lobes, allow us to reasonably exclude a backflow origin (see a similar case in \citealt{2002MNRAS.330..609D}). These tails were not previously detected in NVSS (see \citealt{2021MNRAS.500.3111B}). However, that survey is shallower than the LOFAR observations presented here (the typical RMS level of NVSS is $\sim$0.5 mJy/beam, four times larger than the one in our LOFAR 20$''$ resolution image), and with a poorer short-baselines coverage (thus a lower sensitivity to extended structures). Moreover, the expected peak flux density at the NVSS frequency (1.4 GHz), assuming a typical spectral index $\alpha=-1.5$ for relic plasma, would be a factor 30 lower than the one in our LOFAR images (decreasing from $\sim$1.5 mJy/beam to $\sim$0.05 mJy/beam), laying an order of magnitude below the NVSS noise level. We could estimate an upper limit of $\alpha_{0.14}^{1.44}<-0.52$ and $\alpha_{0.14}^{1.44}<-0.75$ for H1 and H2, respectively, given an RMS noise of 0.5 mJy/beam in the corresponding regions of NVSS images. Further tailored observations, covering the MHz to GHz frequency range, could help in determining the age of these structures. 

In case the hypothesis of a remnant radio structure is confirmed, this could be a genuine case of X-shaped radio galaxy, with a change in the radio axis orientation of almost 40 degrees. This class of sources are among the best candidate to host a binary supermassive black holes system (BSBH, \citealt{2018ApJ...852...48S}). Such a link between empirical evidences of bending jets and the possibility of a geodetic precession in a binary system of AGN, was already proposed decades ago \citep{1980Natur.287..307B}. Recently, 3D hydrodynamic simulations of kpc-scale radio galaxies with a precessing jet \citep{2020MNRAS.499.5765H} allowed to test the formation of the peculiar morphologies among these objects (e.g. X-shaped and S-shaped, see \citealt{2019MNRAS.482..240K} for a summary). Following the work of those authors, the analogous case of J1153.9+5848 could be attributable to a jet with a precessing angle of $\sim$45 degrees, that would form an X-shaped morphology on a Myr timescale with two axes of comparable length. Those authors explain such a jet precession in terms of a BSBH system, formed during a merger with another galaxy. However, alternative explanations have been proposed in the literature, including jet axis realignement \citep{1998ApJ...506L..97N}, and a precessing accretion disk \citep{2008ApJ...674...97K}. High resolution (pc-scale) radio observations could help to disentangle the possible scenarios for this source: in case two radio-emitting black holes are present in its core, they could be detected as flat-spectrum components, analogously to the case of NGC\,7674 \citep{2017NatAs...1..727K}. 

%%%%%%%%%%%%%%%%%%%%%%%%%%%%%%%%%%%%%%%%%%%%%%%%%%%%%%%%%%%%%%%%%%%

\begin{figure*}
    \includegraphics[width=\textwidth]{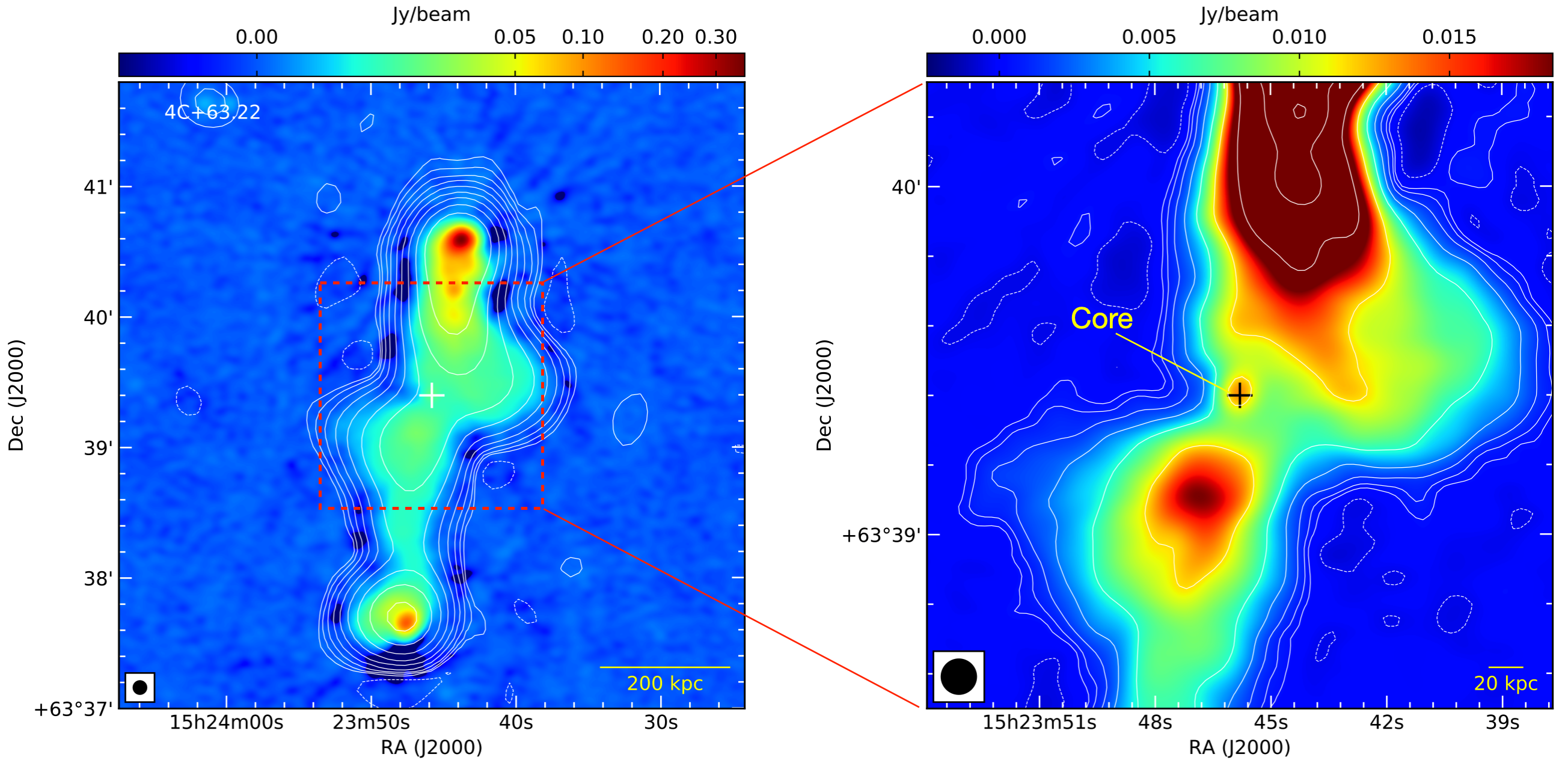}
    \caption{LOFAR images of sources 4C+63.22. Left panel: the color scale represents the 6$''$ resolution image, while contours are from the 20$''$ one; right panel: zoom on the core region, both color scale and contours are drawn from the 6$''$ resolution image. Contours are multiples of the image RMS, namely 3$\times$RMS$\times$(--1, 1, 2, 4, 8, 16, 32, 64, 128, 256). Negative contours are dashed. The cross indicates the X-ray position. The 6$''$ beam is shown in the lower-left corner.}
    \label{fig:4C+63}
\end{figure*}

%%%%%%%%%%%%%%%%%%%%%%%%%%%%%%%%%%%%%%%%%%%%%%%%%%%%%%%%%%%%%%%%%%%

\subsection*{J1238.4+5349}

This source was included in the GRG catalogue compiled from the first data release of LoTSS \citep{2020A&A...635A...5D}. The full resolution image (Fig. \ref{fig:4maps}, bottom-left image) shows two lobes in E-W direction and two compact components in the middle. On the one hand, the easternmost among the latter ones corresponds to the X-ray position from our previous works, and shows a counterpart in the Pan-STARRS i-band image, and is thus coincident with the host galaxy of our GRG (see Fig. \ref{fig:4maps}, bottom-right panel).
On the other hand, the Western component of the inner couple does not correspond to any known source in the SIMBAD astronomical database \citep{2000A&AS..143....9W}, and could thus be interpreted as a knot along the jet. Finally, in the 6 arcsec resolution image from LOFAR a curved, symmetric, emission is visible between the lobes and the core (from the Eastern lobe towards North, and from the Western lobe towards South), resembling the typical morphology of a backflow.      

%%%%%%%%%%%%%%%%%%%%%%%%%%%%%%%%%%%%%%%%%%%%%%%%%%%%%%%%%%%%%%%%%%%

\begin{figure*}
\centering
  \includegraphics[width=15cm]{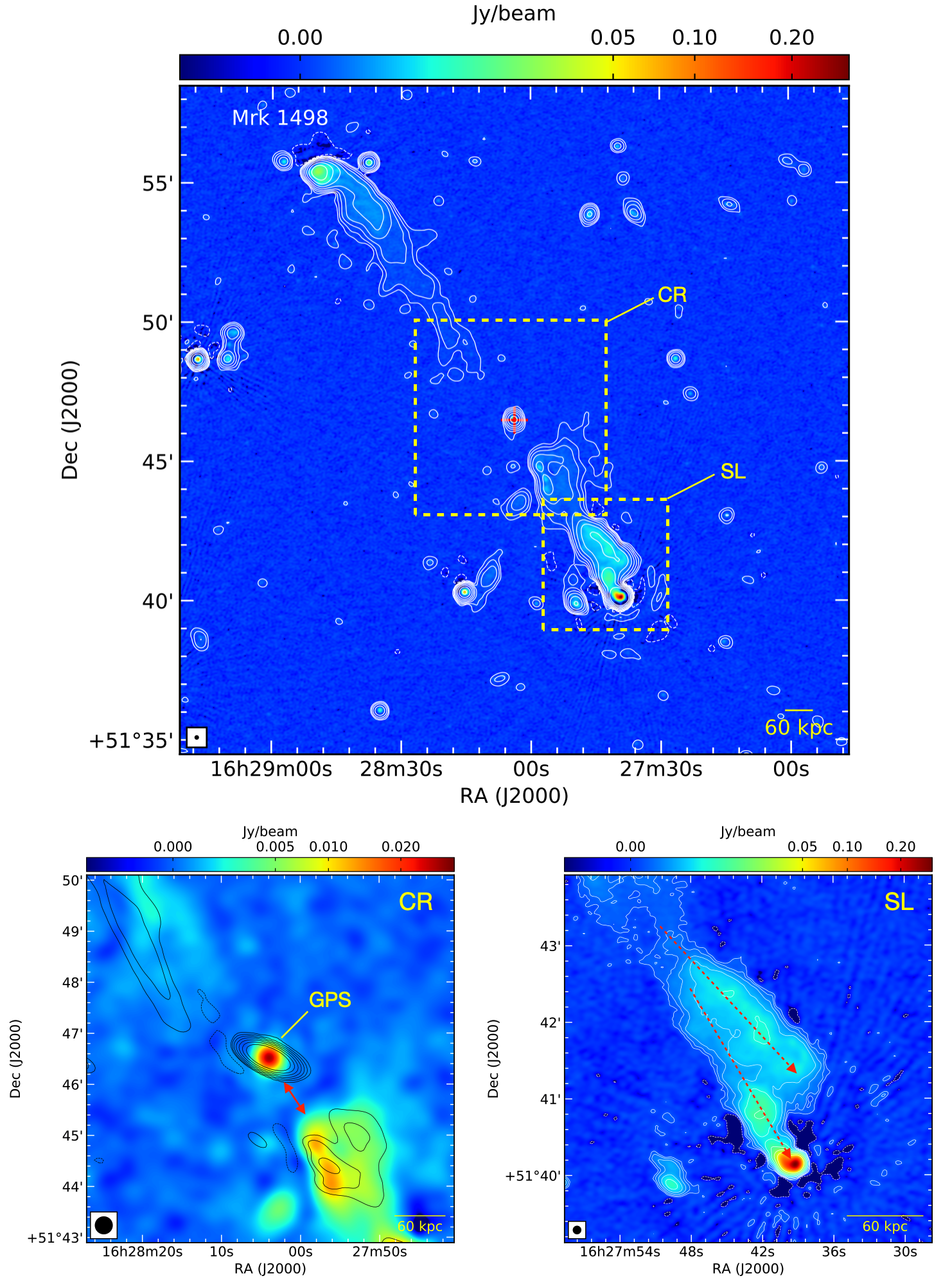}
\caption{LOFAR images of source Mrk\,1498. Top panel: the color scale represents the 6$''$ resolution image, while contours are from the 20$''$ one. The 6$''$ beam is shown in the lower-left corner.  The cross indicates the X-ray position. Bottom-left panel: overplot of the VLA contours at 5 GHz from archival observations (35$''\times$15$''$ resolution) on the LOFAR 20$''$ resolution image. The emission gap of $\sim$40 kpc between the core and the edge of the Southern emission region, visible with both instruments, is marked in red. The 20$''$ LOFAR beam is shown in the lower-left corner. Bottom-right panel: zoom of the LOFAR image at 6$''$ resolution on the south-western lobe, with the two branches marked with a red dashed line. The 6$''$ beam is shown in the lower-left corner. For all images, contours are multiples of the image RMS, namely 3$\times$RMS$\times$(--1, 1, 2, 4, 8, 16, 32, 64, 128). Negative contours are dashed.}
\label{fig:Mrk1498}
\end{figure*}

%%%%%%%%%%%%%%%%%%%%%%%%%%%%%%%%%%%%%%%%%%%%%%%%%%%%%%%%%%%%%%%%%%%

\begin{figure*}
    \includegraphics[width=\textwidth]{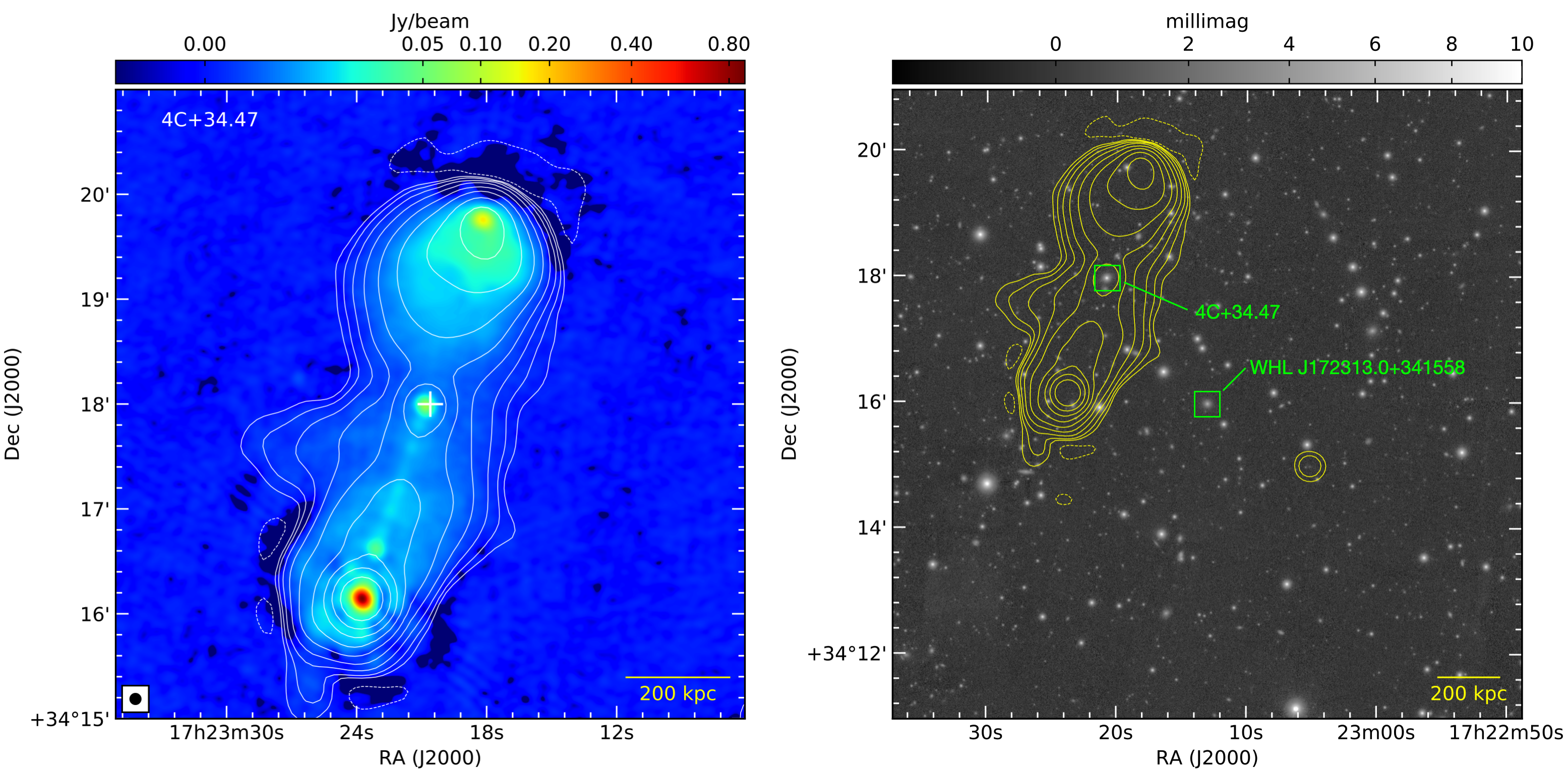}
    \caption{Left panel: LOFAR image of sources 4C+34.47. The color scale represents the 6$''$ resolution image, while contours are from the 20$''$ one. The LOFAR 6$''$ beam is shown in the lower-left corner.  The cross indicates the X-ray position. Right panel: overlay of the LOFAR 20$''$ resolution contours with the Pan-STARRS i-band image centered on the brightest cluster galaxy WHL\,J172313.0+341558. 
    Contours are multiples of the image RMS, namely 3$\times$RMS$\times$(--1, 1, 2, 4, 8, 16, 32, 64, 128, 256). Negative contours are dashed.}
    \label{fig:4C+34}
\end{figure*}

%%%%%%%%%%%%%%%%%%%%%%%%%%%%%%%%%%%%%%%%%%%%%%%%%%%%%%%%%%%%%%%%%%%

\subsection*{J1503.7+6850 (4C+69.18)}

An asymmetric emission is visible for this source (Fig. \ref{fig:4maps}, top-right panel), with the Northern one being brighter than the Southern one. \cite{2012MNRAS.426..851K} discussed how this source almost lies on the plane of the sky, with an inclination angle of 81$^\circ$ with respect to line of sight. This could suggest that the difference in jets brightness could be due to a denser medium probed by the Northern jet (see source Mrk\,1498 below for a similar case). An additional emission spot (indicated as A in the figure) is visible southward of the core: it corresponds to an unidentified optical source in Pan-STARRS, and should thus not be considered as part of the GRG emission. At the lower resolution of NVSS it is not resolved from the core. \cite{2001A&A...370..409L} presented a VLA image of this source at 1.4 GHz, and 10$''$ resolution: the same morphological properties are visible, including component A, that shows a peak flux density of about 6 mJy/beam, resulting in a spectral index $\alpha_{0.14}^{1.44}\sim-0.45$. No signs of restarted radio activity are evident in the LOFAR image or previous ones from the literature, showing a typical FRII morphology.  

%%%%%%%%%%%%%%%%%%%%%%%%%%%%%%%%%%%%%%%%%%%%%%%%%%%%%%%%%%%%%%%%%%%

\subsection*{4C+63.22}
This source was presented as a tentative X-shaped radio galaxy in \cite{2020MNRAS.494..902B}, based on the TIFR GMRT Sky Survey (TGSS, \citealt{2017A&A...598A..78I}) image at 150 MHz, which shows hints of emission distributed in almost perpendicular direction with respect to the main jet axis. The AGN X-ray position nicely corresponds with the radio core. The nature of the diffuse off-axis emission between the lobes and the core was not clearly established at the resolution of the available image (25 arcsec). The LOFAR image presented here (Fig. \ref{fig:4C+63}) offers an improvement in angular resolution by a factor of $\sim$4, and a factor of $\sim$10 in sensitivity, confirming the X-shaped morphology. Nonetheless, we cannot easily classify the off-axis emission as either remnant lobes from a previous radio phase, or plasma backflow created by the present jets. Indeed, the Southern region looks like plasma expanding in the direction from the core region towards to external medium, while the Northern one presents the typical backflow morphology, with a broadening from the lobe towards the core. Finally, the optical images available from surveys are not of sufficient resolution to measure the major axis orientation, and test the scenario proposed by \cite{2016A&A...587A..25G} previously invoked for source J0318+684. Further observations, aiming at studying the plasma aging across these structures, will allow us to shed light on the evolution of this source. 

%%%%%%%%%%%%%%%%%%%%%%%%%%%%%%%%%%%%%%%%%%%%%%%%%%%%%%%%%%%%%%%%%%%

\subsection*{Mrk\,1498}

This source was the subject of a multi-wavelength study by \cite{2019MNRAS.489.4049H}, who discussed a possible reactivation of the nuclear activity following a merger event. Indeed, the source presents an obscured AGN, with a GPS in its core region \citep{2019ApJ...875...88B}. \cite{2020MNRAS.494..902B} discussed the morphological properties visible in the TGSS map at 150 MHz (25$''$ resolution), where basically the core and lobes where detected, while the jet connecting them was not present. The LOFAR image (Fig. \ref{fig:Mrk1498}, top panel) shows for the first time details about the complex structure along the jet, with a bent Northern jet, and a branched Southern lobe, with several knots (see a zoom on this region in figure \ref{fig:Mrk1498}, bottom-right panel). A hint of the two Southern jet branches was already visible in the TGSS image, although the lower resolution did not allow to fully distinguish them. The overall source morphology is asymmetric, with the Northern structure having almost twice the extension of the Southern one (11 arcmin versus 7.5 arcmin, from core to hotspot). A similar asymmetry was discussed by \cite{2019A&A...628A..69S} for 3C\,236, explaining it as a difference in the IGM probed by the two jets. This hypothesis may hold also for Mrk\,1498, moreover it could explain the enhanced surface brightness and forked morphology of the Southern jet and lobe, as a result of a strong interaction with a dense medium. 

Thanks to the enhanced sensitivity of these LOFAR images, it is possible to highlight an emission gap between the well-defined core, and the inner edge of the Southern jet structure (see Fig. \ref{fig:Mrk1498}, bottom-left panel). A similar gap is visible between the core and the edge of Northern lobe, although the latter is less defined if compared with the previous one. On the Southern side, we can measure a separation of about 40 arcsec (corresponding to 38 kpc at the redshift of the source) between the core and the first edge of Southern emission region. A similar gap is visible in archival VLA observations at 5 GHz (35$''\times$15$''$ resolution, see figure \ref{fig:Mrk1498}, bottom-left panel). This is likely due to a collimated jet, that in the region near to the core emits most of its radiation in the propagation direction, and with small energetic losses due to interaction with the medium, making it almost undetectable when lying on the plane of the sky. However, the presence of a GPS in the core of this GRG - as also confirmed by the symmetric morphology visible from archival radio images at pc-scale resolution \citep{2019MNRAS.489.4049H} - seems to suggest that the radio emission in the core region could be from a phase disconnected from the one responsible of the Mpc-scale structure.

%%%%%%%%%%%%%%%%%%%%%%%%%%%%%%%%%%%%%%%%%%%%%%%%%%%%%%%%%%%%%%%%%%%

\subsection*{4C+34.47}
This source presents a well defined core in the LOFAR image (Fig. \ref{fig:4C+34}, left panel), previously classified as a GPS \citep{2019ApJ...875...88B}. Both Southern and Northern lobes have bright and well defined hotspots, sign of an active fueling from the jets. The Northern jet is not visible, probably due to a Doppler dimming effect, while the Southern jet shows a knotty structure between the core and the lobe. These properties were not revealed at the lower resolution of the TGSS image discussed in \cite{2020MNRAS.494..902B}. \cite{2010A&A...523A...9H} first revealed the knotty structure of the Southern jet from VLA observations at 1.4 GHz and 5 GHz, and also provided an estimate of the jet angle to line of sight. On the basis of the jet/counter-jet flux density ratio, and the jet bulk velocity, they infer a jet axis angle within $57^\circ$ from the line of sight. A knotty, discontinuous, jet could be the result of a intermittent activity of the radio core. Another option could be an alternating recollimation/expansion of the jet, resulting in an enhanced flux for the collimated components. Alternatively, a helical jet could produce a similar morphology, with brighter knots corresponding to regions where the jet orientation is close to the line of sight. In appendix \ref{Appendix:chandra}, we present an archival {\it Chandra} image in the 0.5 - 7 keV energy range, were the Southern hotspot and a knot are detected.

Finally, we notice that this GRG can be associated with the galaxy cluster WHL J172313.0+341558 from the \cite{2012ApJS..199...34W} catalogue (see Sec. \ref{cluster}).  Figure \ref{fig:4C+34} (right panel), shows the position of the GRG relative to the Brightest Cluster Galaxy (BCG), having a $z=0.1955$.

%%%%%%%%%%%%%%%%%%%%%%%%%%%%%%%%%%%%%%%%%%%%%%%%%%%%%%%%%%%%%%%%%%%
%%%%%%%%%%%%%%%%%%%%%%%%%%%%%%%%%%%%%%%%%%%%%%%%%%%%%%%%%%%%%%%%%%%

\begin{figure*}
\centering
  \includegraphics[width=\columnwidth]{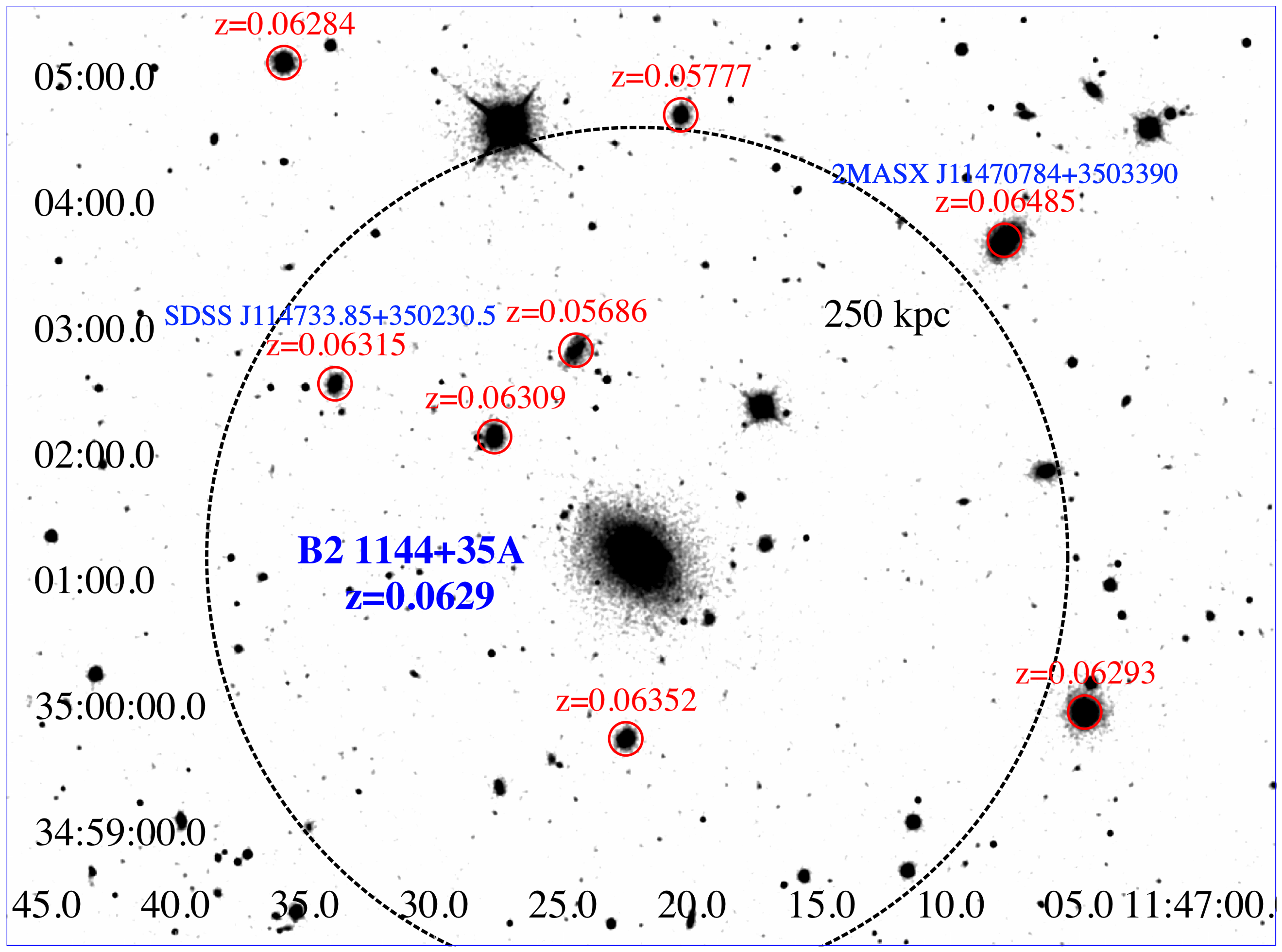}
  \includegraphics[width=\columnwidth]{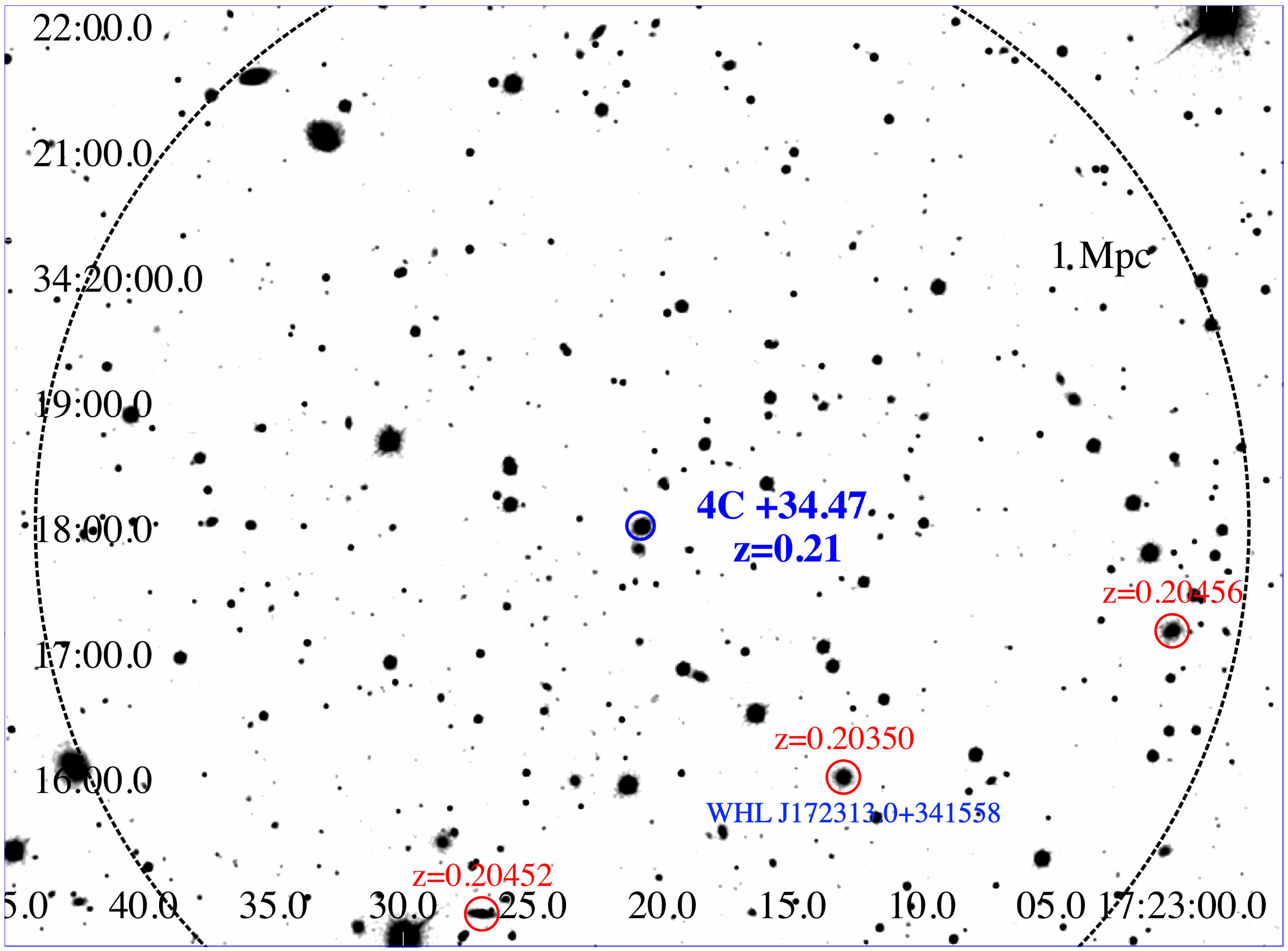}
\caption{\emph{Left panel}: the SDSS R band image of the inner field around B2\,1144+35B \citep{2012ApJS..203...21A}. The blue circle marks the position of the central radio galaxy while the larger dashed black circle corresponds to a distance of 250 kpc computed at its redshift. Red circles mark the location of cosmological neighbors, i.e., optical galaxies having a redshift difference, with respect to that of the central radio galaxy, less than 0.005 (see \citealt{2019ApJS..240...20M,2020ApJ...900L..34M} for more details) as well as that of SDSS J114720.45+350438.5 at $z=$0.05777 lying at $\sim$250 kpc and SDSS J114724.51+350246.5 with a redshift estimate of 0.05686 at $\sim$120 kpc distance, that could belong to the same cosmological structure. B2\,1144+35B appears to be the BCG of a galaxy group/cluster. Finally we highlight that the closest companion galaxy, SDSS J114722.74+350108.8 at $z=$0.06272, located at only $\sim$7 kpc projected distance is not marked here. \emph{Right panel}: The SDSS R band image of the field around 4C\,+34.47 \citep{2012ApJS..203...21A}. The blue circle marks the position of the host elliptical galaxy of 4C\,+34.47 while the larger dashed black circle corresponds to a distance of 1 Mpc computed at the same redshift of the central radio galaxy (i.e., $z=$0.21) and centered on its position. Finally, red circles mark the location of three galaxies. lying within 1 Mpc, and having a redshift difference, with respect to that of the central radio galaxy, less than 0.005 (known as cosmological neighbors; see \citealt{2019ApJS..240...20M, 2020ApJ...900L..34M}). Their presence clearly indicate that 4C+34.47 belong to a galaxy group/cluster.}
\label{fig:clustering}
\end{figure*}

%%%%%%%%%%%%%%%%%%%%%%%%%%%%%%%%%%%%%%%%%%%%%%%%%%%%%%%%%%%%%%%%%%%

\section{Environment analysis}
\label{cluster}

In order to evaluate the influence of environment on the morphology of these GRG, we adopted the same strategy of recent works from the LoTSS collaboration \citep{2019A&A...622A..10C,2020A&A...635A...5D} to investigate the presence of galaxy clusters potentially associated with these GRGs. In particular, we considered the following galaxy clusters catalogues: the Gaussian Mixture Brightest Cluster Galaxy (GMBCG) catalogue \cite{2010ApJS..191..254H}, including 55\,880 clusters from the SDSS DR7 in the redshift range 0.1$<z<$0.55, the  \cite{2012ApJS..199...34W} catalogue, consisting of 132\,684 clusters drawn from SDSS DR8 in the redshift range 0.05$<z<$0.8, and the RedMaPPer catalogue \citep{2014ApJ...785..104R}, also based on SDSS DR8, and providing more than 25\,000 clusters in the redshift range 0.08$<z<$0.55. In addition to those works, we considered the \cite{2012MNRAS.422...25S} catalogue, presenting 10\,443 groups and clusters from SDSS DR7. 

Six out of 9 GRGs studied here, while 9 among the 23 constituting the whole HXGRG sample, are included in the footprint of the mentioned works, not allowing statistical considerations on the full sample. Analogously to \cite{2019A&A...622A..10C}, we considered as criteria for the clusters matching a physical distance $\Delta D\le1$ Mpc at the redshift of our objects, and a maximum difference in redshift $\Delta z\le0.01$. In this way, we found an association for 2 sources out of 6: B2\,1144+35B, matching with one of the clusters from the \cite{2012MNRAS.422...25S} catalogue (identified with ID 2259 in that work), and 4C+34.47, associated with the galaxy cluster WHL J172313.0+341558 from the \cite{2012ApJS..199...34W} catalogue. Both sources are included in the sub-sample of 9 GRGs with LoTSS data presented here. The remaining 4 sources presented here that do not show a match with galaxy clusters are Mrk\,1498, J0801.7+4764, J1153.9+5848, and J1238.4+5349.

To confirm this result, we performed an environment analysis adopting the same methods and procedures outlined in \cite{2019ApJS..240...20M,2020ApJS..247...71M,2020ApJ...900L..34M}. We searched for cosmological neighbors in the SDSS lying within the 2\,Mpc radius computed at the same redshift of the central radio galaxy, with all SDSS magnitude flags indicating a galaxy-type object, and having a maximum spectroscopic redshift difference $\Delta z\le0.005$ with respect to that of the central target, corresponding to the maximum velocity dispersion in groups and clusters of galaxies (see e.g. \citealt{1993MNRAS.261..827M,2004MNRAS.348..866E,2006ApJS..167....1B}). We could confirm a cosmological neighbors over-densities around B2\,1144+35B and 4C\,+34.47, in agreement with the results from the cross-match of the mentioned catalogues.

In the case of B2\,1144+35B, we found 9 cosmological neighbors in addition to SDSS J114720.45+350438.5 at $z$=0.05777 lying at $\sim$250 kpc and SDSS J114724.51+350246.5 with a redshift estimate of 0.05686 at $\sim$120 kpc distance (see Figure \ref{fig:clustering}, left panel). In particular 3 cosmological neighbors are located within a circular region of 250 kpc, one of them showing a prominent $H\alpha$ in its optical spectra: SDSS J114733.85+350230.5, at $z=0.06315$  as occurs for SDSS J114702.75+350720.4 ($z=$0.06298) at larger distance. A very close companion galaxy, namely: SDSS J114722.74+350108.8 at $z=$0.06272 is also located at only $\sim$7 kpc projected distance (this is not highlighted in Fig. \ref{fig:clustering}). The probability to have this number of cosmological neighbors at $z\sim$0.06 is less than a few percent, as shown in \cite{2019ApJS..240...20M}, thus strongly suggesting the presence of a galaxy cluster. Finally, we highlight that B2\,1144+35B is most probably the BGC of this galaxy cluster being located at $\sim$70 kpc from the centroid of all positions of all other cosmological neighbors. The distribution of the central region (i.e., within 250 kpc) of all SDSS optical sources, marked with their redshift estimate if present, is shown in Fig. \ref{fig:clustering}.

For 4C\,+34.47, we found 3 cosmological neighbors within 1\,Mpc distance from the position of the central radio galaxy, all marked with red circles in Fig. \ref{fig:clustering}, right panel. A fourth galaxy at $z$=0.20479 lies at 1.12 Mpc in the north-western direction, being probably part of the same cosmological structure. With respect to B2\,1144+35B, 4C\,+34.47 lies at higher redshift and given the number of cosmological neighbors we can certainly claim that also this radio galaxy is harbored in a galaxy rich environment (i.e., a group or a cluster of galaxies).

%%%%%%%%%%%%%%%%%%%%%%%%%%%%%%%%%%%%%%%%%%%%%%%%%%%%%%%%%%%%%%%%%%%
%%%%%%%%%%%%%%%%%%%%%%%%%%%%%%%%%%%%%%%%%%%%%%%%%%%%%%%%%%%%%%%%%%%

\section{Comparison with radio-selected GRG from LoTSS DR1}

\begin{figure}
    \includegraphics[width=\columnwidth]{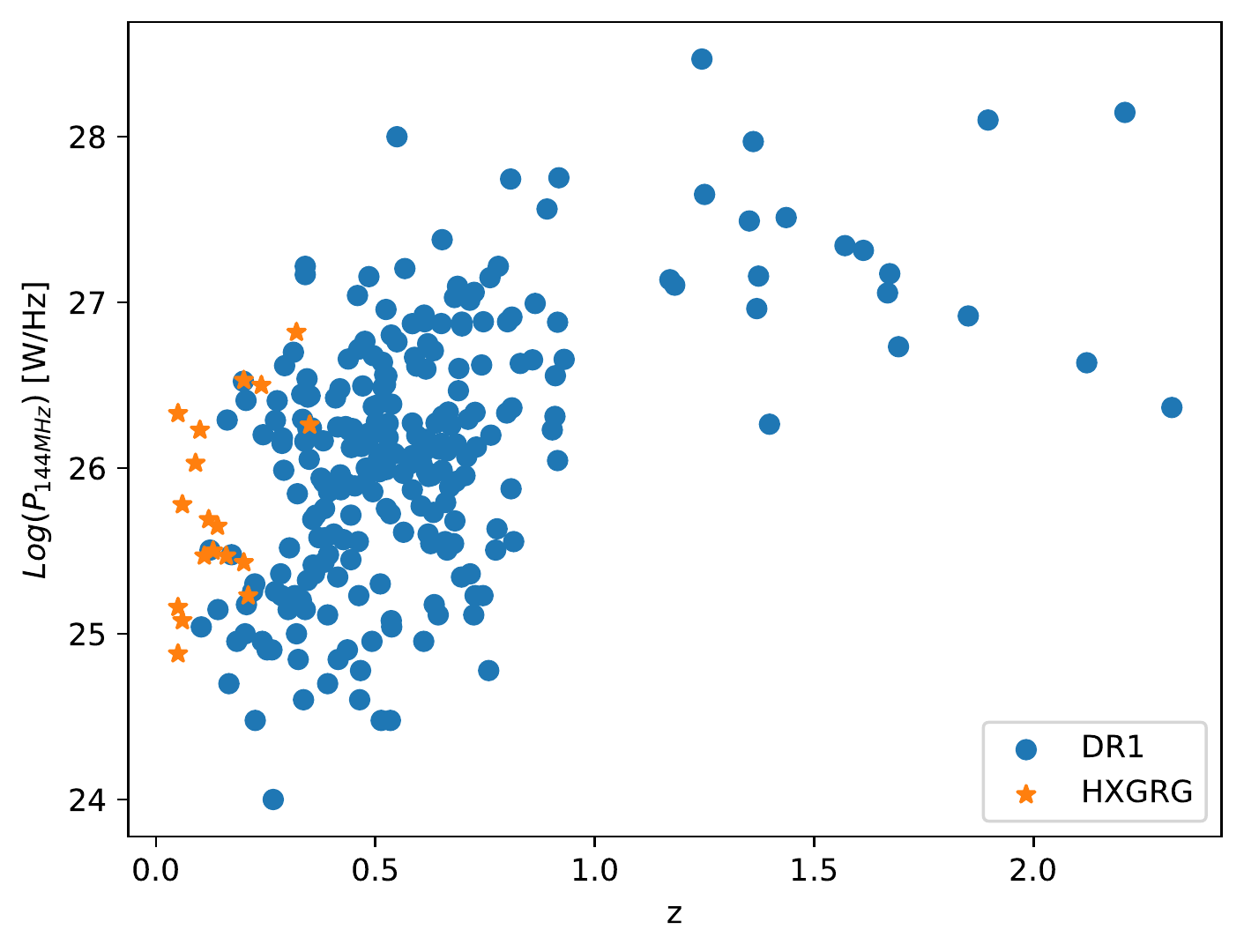}
    \includegraphics[width=\columnwidth]{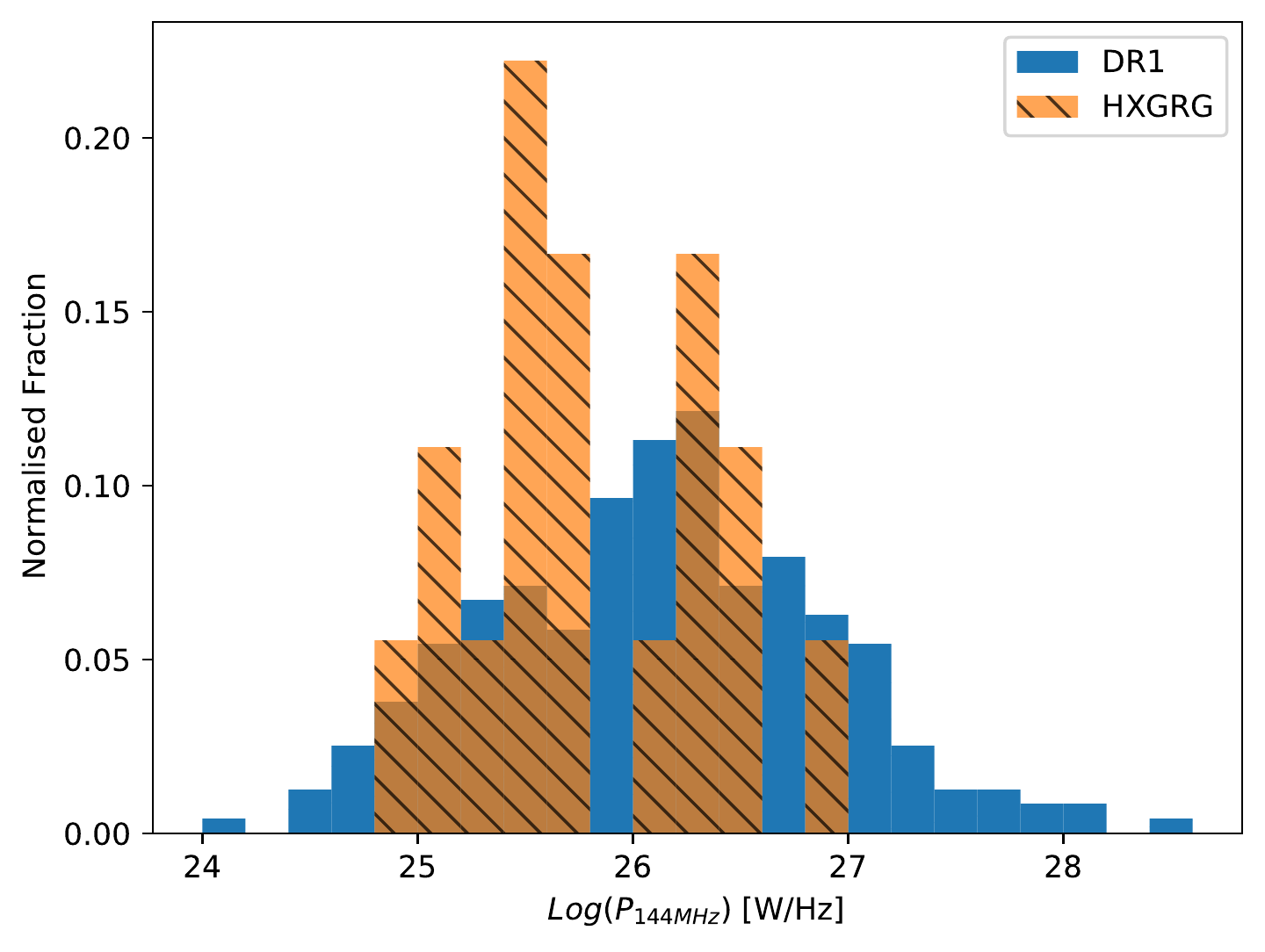}
    \includegraphics[width=\columnwidth]{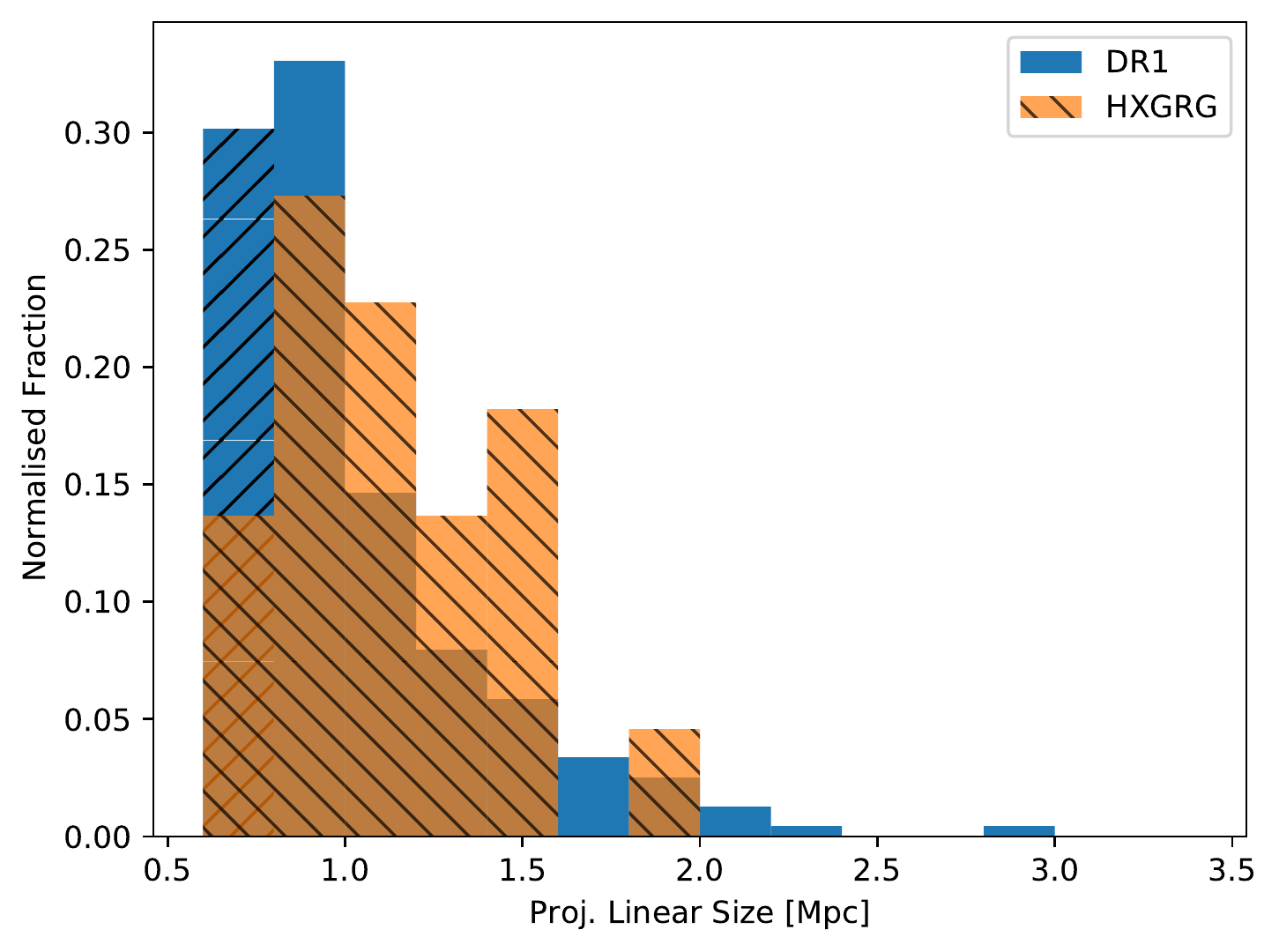}
    \caption{Comparison of the HXGRG sample with the radio-selected GRG from LoTSS DR1 (DR1, \citealt{2020A&A...635A...5D}). Top panel: redshift vs total radio power at 144 MHz; middle panel: total radio power at 144 MHz; bottom panel: projected linear sizes.}
    \label{fig:comparison}
\end{figure}

The first selection step of our sample, i.e. the detection in hard X-ray catalogues, introduces a bias towards the high accretion regime in the core of these GRG. Indeed, the Eddington ratios of our sources fall in the range 0.01--0.37 \citep{2018MNRAS.481.4250U, 2021MNRAS.500.3111B}. In principle, this could play a role on the evolution of their radio phase, with respect to traditional radio-selected GRG samples. While a dedicated work on the detailed comparison with a control sample of radio-selected GRG is under way, we can here compare the main properties of our HXGRG sample with the 239 radio-selected ones from LoTSS DR1 \citep{2020A&A...635A...5D}. In figure \ref{fig:comparison}, we compare the redshift vs radio power at 144 MHz for the two samples (top panel), the radio power distributions (middle panel, normalized fractions), and the projected linear sizes distributions (bottom panel, normalized fractions). In addition to the LoTSS DR2 data presented in this work, we also considered the TGSS data (in the same band and at a similar resolution of 25 arcsec) to improve the statistics of our HXGRG sample. In this way, we could collect the total radio powers at $\sim$150 GHz for 18 over 23 objects of the sample, namely all HXGRG north of --53$^\circ$ Dec except the GRG candidate SWIFT\,J0225.8+5946, for which a redshift estimate is still not available. The estimated values from TGSS, for sources not included in LoTSS DR2, are given in table \ref{tab:TGSS}. For the redshift and linear size, we considered the values presented in \cite{2020MNRAS.494..902B} and \cite{2021MNRAS.500.3111B}, available for the full sample except SWIFT\,J0225.8+5946 (22 objects in total).

As visible in Fig. \ref{fig:comparison} (top panel), HXGRG have a redshift $z<0.4$, and lye in the lower envelope of the LoTSS DR1 GRGs. This is likely due to the selection criteria of the hard X-ray AGN sample from which they are drawn. Indeed, the \emph{INTEGRAL}/IBIS and \emph{Swift}/BAT surveys are of relatively low sensitivity, thus preferably detecting brighter and nearby objects (typically with $z$ below 0.2-0.3, see \citealt{2020NewAR..9001545M}). The total radio powers distributions at 144 MHz (Fig. \ref{fig:comparison} middle panel) are comparable for the two samples, with HXGRG spanning the range $10^{25}-10^{27}$ W/Hz, well within the one found for LoTSS DR1 GRGs ($10^{24}-10^{28.5}$ W/Hz). A Kolmogorov-Smirnov (KS) test on the two distributions results in a p-value>10\%, not allowing to discard the null-hypothesis of a common parent distribution for the two. Despite the limited statistics (18 sources), this might indicate that the hard X-ray selection does not imply a difference in total radio power integrated over a Myr-timescale activity, although the Eddington ratios of our GRGs falls in the high tail of the distribution for the radio-selected ones (0.01--0.37, \citealt{2018MNRAS.481.4250U, 2021MNRAS.500.3111B}, to be compared with e.g. figure 8 of \citealt{2020A&A...642A.153D}). The projected linear size distributions are plotted in Fig. \ref{fig:comparison} (bottom panel) for 22 HXGRG vs the full sample of 239 radio-selected GRGs from LoTSS DR1. The HXGRG sample shows, on average, larger values, with a mean value of 1.15 Mpc versus 1.03 Mpc for the LoTSS DR2 GRGs, although the linear size range covered by the two samples is similar. A KS test confirms that the two distributions are distinct, giving a p-value <5\% for the null-hypothesis. Nevertheless, an improved statistics for HXGRG is necessary to asses this conclusion. If confirmed, this result would be consistent with the larger fraction of GRG found by \cite{2016MNRAS.461.3165B} among hard X-ray selected AGN. A similar result was also found by \cite{2020A&A...642A.153D}, when comparing the sizes of high-excitation giant radio galaxies (HEGRGs, i.e. objects with an efficient accretion similarly to our HXGRG) with the ones of low-excitation giant radio galaxies (LEGRGs): on average, HEGRGs reach larger extents than LEGRGs (see Fig. 7 of their work). 

From a morphological point of view, \cite{2020A&A...635A...5D} report a 5\% of double-double GRG. Among the 9 objects presented here, we do not have any double-double, but we can account for two clear cases among the full sample of 23 HXGRG. This gives a fraction of $9^{+11}_{-6}$\%, following Poissonian statistics \citep{1986ApJ...303..336G}, consistent within errors with the double-double fraction for GRGs in LoTSS DR1. \cite{2020A&A...635A...5D} also discuss the FRI/FRII and HyMoRS fractions (the latter being sources with a FRI morphology on one side and FRII on the other side \citealt{1996MNRAS.282..837S}), finding 90\% of FRII, while $\sim$2.5\% of HyMoRS. Regarding our HXGRG sample, all sources except for one (B2 1144+35B) are FRII, resulting in a fraction of $95\pm5$\%, again compatible within errors. No cases of HyMoRS are present (as expected for small samples). Nevertheless, those authors do not provide a statistic for reoriented/precessing jets, X-shaped morphology, or GPS nuclei, so we postpone to a future work a full comparison of the morphological properties and restarted radio activity fraction between radio-selected and hard X-ray selected GRGs.

Finally, we can compare the environment of the two samples, although with a limited statistics for HXGRG. \cite{2020A&A...635A...5D} found that at least $\sim$16\% of their GRG at $z<0.55$ are the BCG of a cluster or group of galaxies. From the analysis of our sample, 2/9 sources are found to be associated with galaxy clusters from catalogues. This gives a fraction of $22^{+30}_{-14}$\%, consistent within errors with what obtained for the LoTSS DR1 GRGs.

Overall, the comparison of radio power, linear size, and redshifts for the two samples suggests that the main effects of the hard X-ray selection are: 1) a cut of the redshift distribution, i.e. favouring nearby objects at $z<0.4$; 2) a slightly larger growth, on average, for HXGRG with respect to radio-selected GRG samples. No statistically significant difference in total radio power distributions of HXGRG with respect to radio-selected GRG is found.

%%%%%%%%%%%%%%%%%%%%%%%%%%%%%%%%%%%%%%%%%%%%%%%%%%%%%%%%%%%%%%%%%%%

\begin{table}
    \centering
    \begin{tabular}{lccccccc}
    \hline
    Source                  &  Total flux density   &  Total radio power\\
                            &  (Jy)                 & Log(W/Hz)         \\
    \hline 
    B3\,0309+411B			& 0.75$\pm$0.08		    & 25.50   \\
    SWIFT\,J0636.5-2036	    & 38.2$\pm$3.8\,\,\,	& 26.33   \\
    PKS\,0707-35			& 8.97$\pm$0.90 	    & 25.47   \\
    4C\,73.08			    & 7.33$\pm$0.73		    & 25.78   \\
    HE\,1434-1600		    & 0.93$\pm$0.09		    & 25.65   \\
    IGR\,J14488-4008		& 1.41$\pm$0.14		    & 25.69   \\
    IGR\,J17488-2338		& 2.06$\pm$0.21		    & 26.50   \\
    4C+74.26				& 6.78$\pm$0.68			& 26.23   \\
    PKS\,2331-240			& 1.24$\pm$0.12			& 24.88   \\
    \hline
    \end{tabular}
    \caption{Total flux density and radio power at 150 MHz for sources from the HXGRG sample with available TGSS data, and not included in LoTSS DR2.}
    \label{tab:TGSS}
\end{table}

%%%%%%%%%%%%%%%%%%%%%%%%%%%%%%%%%%%%%%%%%%%%%%%%%%%%%%%%%%%%%%%%%%%

%%%%%%%%%%%%%%%%%%%%%%%%%%%%%%%%%%%%%%%%%%%%%%%%%%%%%%%%%%%%%%%%%%%
%%%%%%%%%%%%%%%%%%%%%%%%%%%%%%%%%%%%%%%%%%%%%%%%%%%%%%%%%%%%%%%%%%%

\section{Summary and Conclusions}

We presented LOFAR images of 9 GRG selected from hard X-ray catalogues, at an angular resolution of 6 arcsec (typical RMS of 150 $\mu$Jy/beam) and 20 arcsec (typical RMS of 300 $\mu$Jy/beam), realized with LoTSS data. Thanks to the outstanding quality and dynamic range of LOFAR images, we could probe the complex structure of these sources, identifying for the first time diffuse, remnant plasma on the Mpc scale, off-axis emission suggesting jet reorientation/precession, and in one case an inner jet with a different axis with respect to the one connecting the lobes. In particular, for one source (B2 1144+35B), we could recover an extended, faint emission southwards of one of the lobes at a better resolution with respect to previous works \citep{1999ApJ...522..101G}, allowing us to classify this source as a fat-double. With a morphology consistent with a jet realignement, this class of sources are supposed to result from an inner accretion with a perturbed geometry, probably due to a merger. The inner jet axis detected from LOFAR images of this source, showing a different angle with respect to the most ancient and extended emission, could suggest a counter-clock jet reorientation along a Myr time scale. Similarly, another source showed a symmetrical, faint emission on a different axis with respect to the jet one, consistent with relic lobes from a previous radio phase or jet precession (J1153.9+5848). Finally, for 4C+34.47, already showing two radio phases from our previous studies, we could confirm a discontinuous, knotty jet at a frequency an order of magnitude lower than the previous study from \cite{2010A&A...523A...9H}. This is suggestive of either an intermittent radio activity, or of a plasma recollimation/expansion along the jet. 

An environment analysis, performed through a cross match with four among the most recent clusters catalogue \citep{2010ApJS..191..254H, 2012ApJS..199...34W, 2012MNRAS.422...25S, 2014ApJ...785..104R}, allowed us to find an association with two of the presented sources: B2\,1144+35B and 4C+34.47. This was also verified through an independent environment analysis from SDSS images. In this regards, the cluster might have concurred to the fat-double morphology of source B2\,1144+35B, perturbing the jet axis on a Myr timescale. On the contrary, no morphological perturbation seems to be present for 4C+34.47. The remaining 7 sources from the full HXGRG sample falling in the footprint of the mentioned catalogues do not show a match with clusters.

Finally, we could compare the main properties of our full sample of 23 HXGRG with the 239 radio-selected GRGs from LoTSS DR1 \citep{2020A&A...635A...5D}. While the distributions of total radio power at 144 MHz are consistent, a comparison of the linear sizes for the two samples suggests that HXGRG are on average more extended than radio-selected GRG, although with a limited statistics. This is in agreement with the findings of \cite{2016MNRAS.461.3165B} about the parent sample of hard X-ray selected radio galaxies, from which the HXGRG sample is drawn. Indeed, those authors found that the initial selection step would favour discovery of giant-sized radio galaxies, the latter being almost four times more abundant than in radio-selected radio galaxies samples. A comparison of the environments for the HXGRG and LoTSS DR1 GRG samples gave similar results for the two samples, with a comparable fraction of GRG matching with galaxy clusters or groups.

These results stress how the high resolution and sensitivity at the low frequency radio regime offered by LOFAR and the LoTSS survey in particular, are key to the understanding of the Myr time scale evolution of radio galaxies, and their radio duty cycle. Ultimately, this could give significant inputs on the timescale of the radio-mode feedback on the host galaxy \cite{2013Sci...341.1082M}. In the context of the GRACE project, we are carrying out an European VLBI Network (EVN) large project, which high-resolution (pc-scale) data will shed light on the inner jet properties of these sources, particularly the ones showing a jet precession or reorientation.

%%%%%%%%%%%%%%%%%%%%%%%%%%%%%%%%%%%%%%%%%%%%%%%%%%%%%%%%%%%%%%%%%%%

\section*{Acknowledgements}

The GRACE project acknowledges financial support by INAF-IAPS, in the `Bando Nuove Idee 2019' framework. 
%%%
GB, FP, LB, AM, and PU acknowledge financial support under the INTEGRAL ASI-INAF agreement 2019-35-HH.0.
%%%
MB acknowledges support from the ERC-Stg DRANOEL, no 714245.
%%%
GB and FP acknowledge support from a grant PRIN-INAF SKA-CTA 2016.
%%%
AB acknowledges support from the VIDI research programme with project number 639.042.729, which is financed by the Netherlands Organisation for Scientific Research (NWO).
%%%
AD acknowledges support by the BMBF Verbundforschung under the grant 05A20STA. 
%%%
RJvW acknowledges support from the ERC Starting Grant ClusterWeb 804208.
%%%
This work is also supported by the Physics Department of Turin University (UniTO) that joined the LOFAR-IT consortium thanks to the collaboration with the Consorzio Interuniversitario per la Fisica Spaziale (CIFS) and the ``Departments of Excellence 2018-2022'' Grant awarded by the Italian Ministry of Education, University and Research (MIUR) (L. 232/2016). The analysis is partially based on the OCCAM computing facility hosted by C3S\footnote{\href{http://c3s.unito.it/}{http://c3s.unito.it/}} at UniTO \citep{17:occam:chep}. This research has made use of resources provided by the Ministry of Education, Universities and Research for the grant MASF\_FFABR\_17\_01. This investigation is supported by the National Aeronautics and Space Administration (NASA) grants GO9-20083X and GO0-21110X.
%%%	
LOFAR \citep{2013A&A...556A...2V} is the LOw Frequency ARray designed and constructed by ASTRON. It has observing, data processing, and data storage facilities in several countries, which are owned by various parties (each with their own funding sources), and are collectively operated by the ILT foundation under a joint scientific policy. The ILT resources have benefitted from the following recent major funding sources: CNRS-INSU, Observatoire de Paris and Universit\'e d'Orl\'eans, France; BMBF, MIWF-NRW, MPG, Germany; Science Foundation Ireland (SFI), Department of Business, Enterprise and Innovation (DBEI), Ireland; NWO, The Netherlands; The Science and Technology Facilities Council, UK; Ministry of Science and Higher Education, Poland; Istituto Nazionale di Astrofisica (INAF), Italy. 
%%%
This research made use of the Dutch national e-infrastructure with support of the SURF Cooperative (e-infra 180169) and the LOFAR e-infra group. The J\"ulich LOFAR Long Term Archive and the GermanLOFAR network are both coordinated and operated by the J\"ulich Supercomputing Centre (JSC), and computing resources on the supercomputer JUWELS at JSC were provided 15 by the Gauss Centre for Supercomputinge.V. (grant CHTB00) through the John von Neumann Institute for Computing (NIC). This research made use of the University of Hertfordshirehigh-performance computing facility and the LOFAR-UK computing facility located at the University of Hertfordshire and supported by STFC [ST/P000096/1], and of the Italian LOFAR IT computing infrastructure supported and operated by INAF, and by the Physics Department of Turin university (under an agreement with Consorzio Interuniversitario per la Fisica Spaziale) at the C3S Supercomputing Centre, Italy.
%%%
The National Radio Astronomy Observatory is a facility of the National Science Foundation operated under cooperative agreement by Associated Universities, Inc.
%%%
We thank the staff of the GMRT that made these observations possible. GMRT is run by the National Centre for Radio Astrophysics of the Tata Institute of Fundamental Research.
%%%
This research made use of APLpy, an open-source plotting package for Python \citep{2012ascl.soft08017R}.
%%%
This research has made use of the SIMBAD database,
operated at CDS, Strasbourg, France
%%%
The Pan-STARRS1 Surveys (PS1) and the PS1 public science archive have been made possible through contributions by the Institute for Astronomy, the University of Hawaii, the Pan-STARRS Project Office, the Max-Planck Society and its participating institutes, the Max Planck Institute for Astronomy, Heidelberg and the Max Planck Institute for Extraterrestrial Physics, Garching, The Johns Hopkins University, Durham University, the University of Edinburgh, the Queen's University Belfast, the Harvard-Smithsonian Center for Astrophysics, the Las Cumbres Observatory Global Telescope Network Incorporated, the National Central University of Taiwan, the Space Telescope Science Institute, the National Aeronautics and Space Administration under Grant No. NNX08AR22G issued through the Planetary Science Division of the NASA Science Mission Directorate, the National Science Foundation Grant No. AST-1238877, the University of Maryland, Eotvos Lorand University (ELTE), the Los Alamos National Laboratory, and the Gordon and Betty Moore Foundation.

\section*{Data Availability}
The data underlying this article will be shared on reasonable request to the corresponding author.



\bibliographystyle{mnras}
\bibliography{GRG_references} % if your bibtex file is called example.bib

%%%%%%%%%%%%%%%%%%%%%%%%%%%%%%%%%%%%%%%%%%%%%%%%%%
%%%%%%%%%%%%%%%%% APPENDICES %%%%%%%%%%%%%%%%%%%%%
%%%%%%%%%%%%%%%%%%%%%%%%%%%%%%%%%%%%%%%%%%%%%%%%%%

%%%%%%%%%%%%%%%%%%%%%%%%%%%%%%%%%%%%%%%%%%%%%%%%%%%%%%%%%%%%%%%%%%%%
\appendix
%%%%%%%%%%%%%%%%%%%%%%%%%%%%%%%%%%%%%%%%%%%%%%%%%%%%%%%%%%%%%%%%%%%%

\section{The X-ray detection of the radio jet in 4C+34.47}
\label{Appendix:chandra}

\begin{figure*} 
   \includegraphics[width=\textwidth]{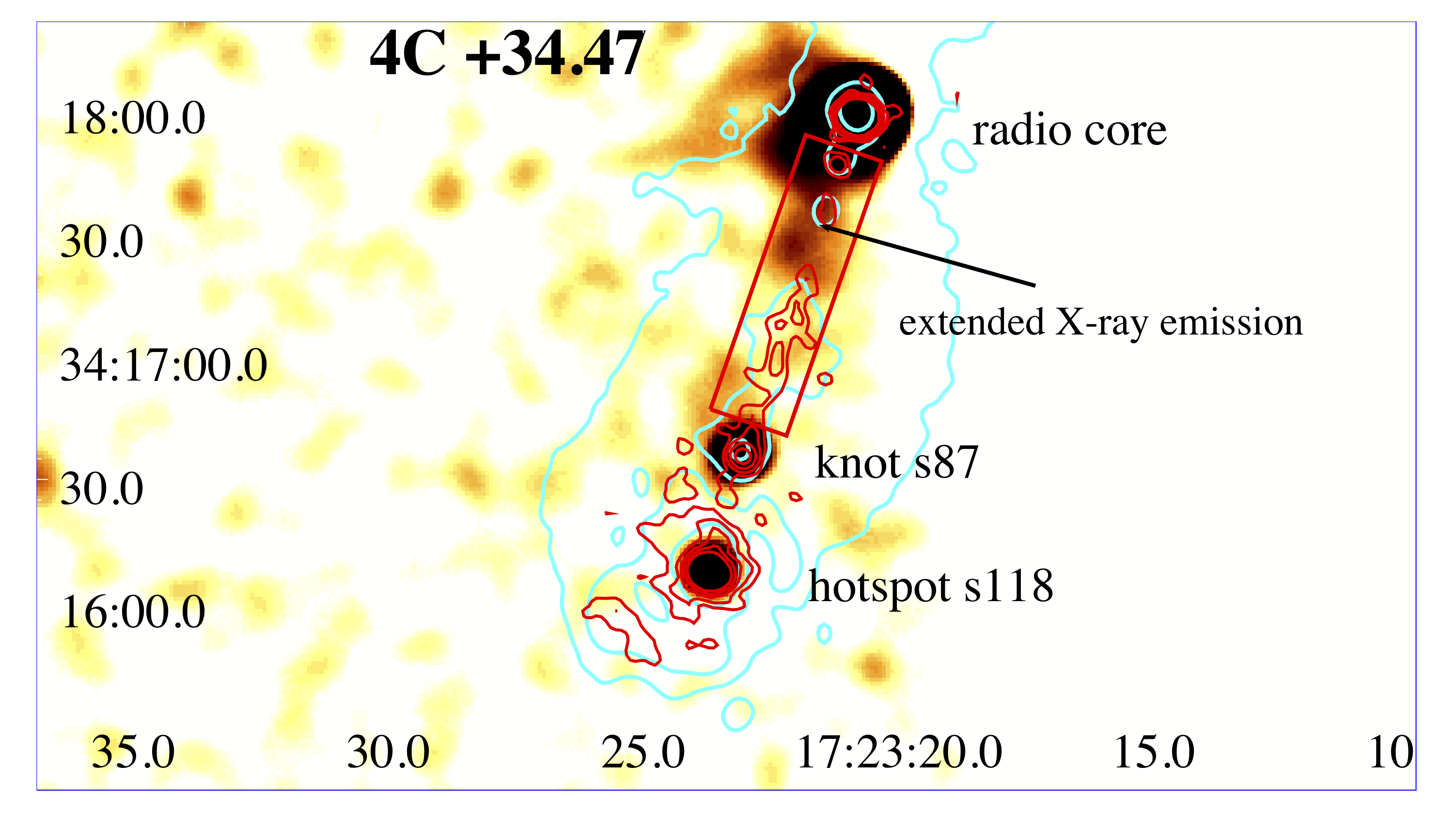}
\caption{{\it Chandra} image in the 0.5 - 7 keV energy range (exposure corrected and with the readout streak removed) of 4C\,+34.47 with 1.4 GHz radio contours overlaid in red and those at 150 MHZ obtained from the LOFAR observation shown in cyan. Low frequency radio contours are drawn starting from the level of 0.002 mJy/beam increasing by a factor of 4, while those at 1.4 GHz start at 0.0001 mJy/beam increasing by a factor of 2 reach 0.032 mJy/beam. The X-ray image is smoothed with a Gaussian kernel of size equal to 7.872\arcsec and pixels have size large as twice native pixel size of Chandra (i.e., 0.984\arcsec). It is quite evident the X-ray detection of the counterpart of both the knot s87 and the hotspot s118 as well as the presence of extended X-ray emission between the position of the radio core of 4C +34.47, lying at the edge of the chip, and the s87 jet knot. All high energy counterparts of these radio structures are detected at level of significance higher than 5$\sigma$ Gaussian equivalent.}
\label{fig:chandra}
\end{figure*}

%%%%%%%%%%%%%%%%%%%%%%%%%%%%%%%%%%%%%%%%%%%%%%%%%%%%%%%%%%%%%%%%%%%%

We searched in the {\it Chandra} archive and discovered that 4C\,+34.47, lying in the field of CSWA 14 (obs. ID 18224), was serendipitously observed in the X-rays on Sept. 2016. We retrieved the dataset from the {\it Chandra} archive, and performed the data reduction and analysis following the same procedures described in the {\it Chandra} Interactive Analysis of Observations (CIAO; \citealt{2006SPIE.6270E..1VF}) threads using CIAO v4.11 and {\it Chandra} Calibration Database v4.8.4. We found that the X-ray counterpart of 4C\,+34.47 lies on the edge of the ACIS-S chip 3, a few arcminutes angular separation from the aim point, while the northern lobe is in the gap between different chips. However, overlaying radio contours to the X-ray image, we discovered the X-ray counterpart of the brightest knot and the hotspot in the southern lobe, at 87\arcsec and 118\arcsec angular separation from the radio core, respectively. These radio components are detected in 0.5 - 7 keV energy range at level of significance greater than 5$\sigma$ Gaussian equivalent.

Adopting the nomenclature proposed in \cite{2011ApJS..197...24M} we labelled the first radio feature as s87 while the hotspot as s118. We measured 70 and 48 X-ray photons in the 0.5-7 keV energy range for s87 and s118, respectively. We also claim the detection of extended X-ray emission along the radio axis, in the southern direction, located between the radio core and the knot s87 with 153 X-ray photons located within the red rectangular region shown in Fig. \ref{fig:chandra}, and correspondent to a detection of $>$5$\sigma$ level of confidence.

In Figure \ref{fig:chandra} we show the {\it Chandra} 0.5-7 keV flux map, exposure corrected and with radio contours overlaid, drawn from the 1.4 GHz radio map retrieved from the Historical archive of the Very Large Array\footnote{\href{http://archive.nrao.edu/nvas/}{http://archive.nrao.edu/nvas/}} and shown together with those at 144 MHz one obtained with LOFAR. The {\it Chandra} flux map has been built adopting the same procedure adopted for the 3CR {\it Chandra} Snapshot Survey (see e.g. \citealt{2010ApJ...714..589M,2013ApJS..209...10M,2015ApJS..220....5M,2018ApJS..235...32S,2020ApJS..250....7J} for additional details) as well as adopting the \textsc{fluximage} task of CIAO. Finally we highlight that the readout streak, present in the {\it Chandra} observation given the relatively high flux of the X-ray nucleus, was also removed in the image shown in Fig. \ref{fig:chandra} according to the standard procedure\footnote{\href{https://cxc.cfa.harvard.edu/ciao/threads/acisreadcorr/}{https://cxc.cfa.harvard.edu/ciao/threads/acisreadcorr/}}.

%%%%%%%%%%%%%%%%%%%%%%%%%%%%%%%%%%%%%%%%%%%%%%%%%%%%%%%%%%%%%%%%%%%%%%%%
%%%%%%%%%%%%%%%%%%%%%%%%%%%%%%%%%%%%%%%%%%%%%%%%%%%%%%%%%%%%%%%%%%%%%%%%

\section{Serendipitous discovery of a radio galaxy}
\label{Appendix:1}

\begin{figure*} 
   \includegraphics[width=\textwidth]{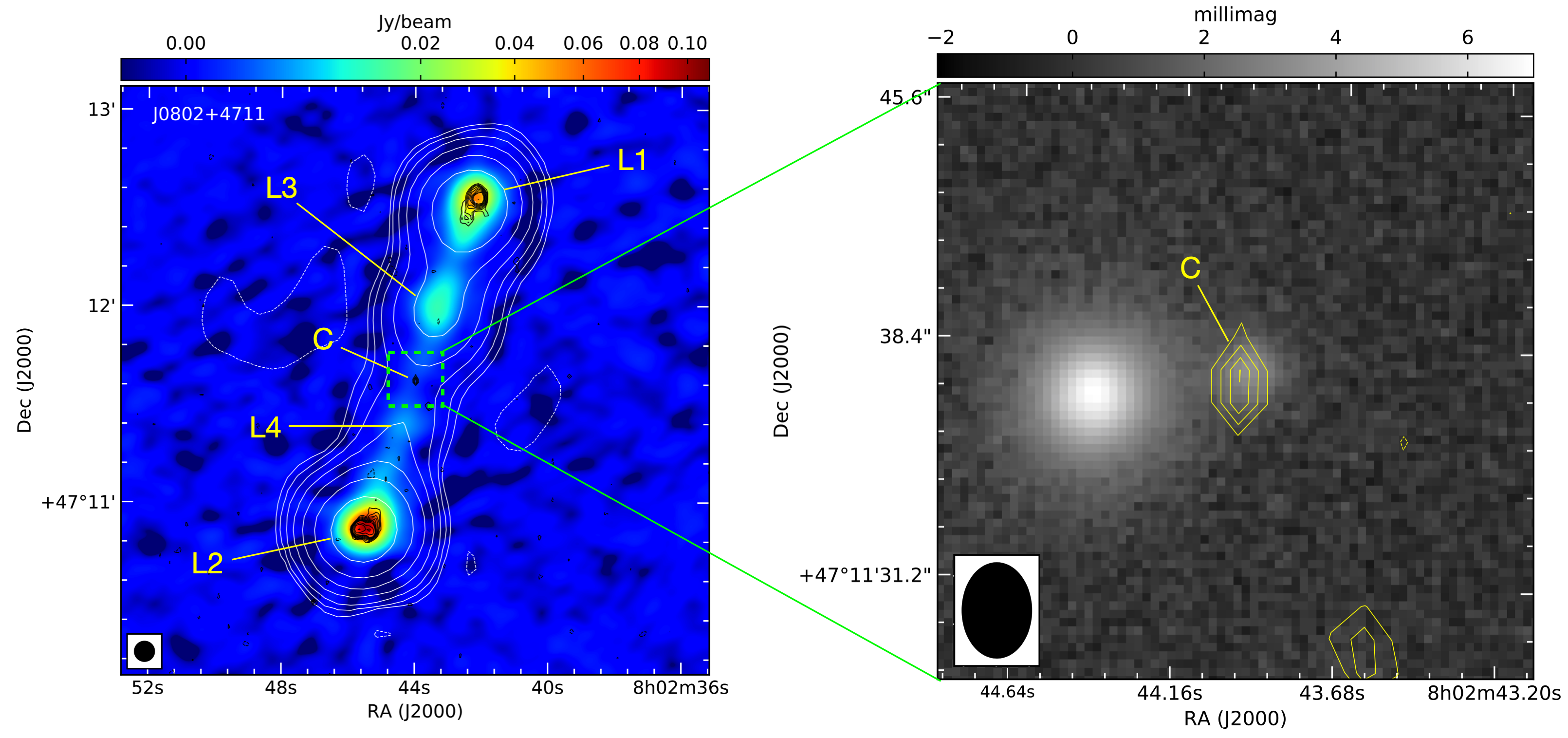} \\
\caption{Left panel: LOFAR image at 6$''$ resolution of the newly discovered RG J0802+4711. Black contours are from VLASS, and correspond to RMS multiples, namely 0.16$\times$(-3, 3, 4, 5, 6, 7, 8, 9, 10, 15, 20, 30, 40, 50, 100). White contours are from the LOFAR image at 20$''$ resolution, and corresponds to 3$\sigma$ multiples, namely 3$\sigma\times$(-1, 1, 2, 4, 8, 16, 32, 64, 128, 256). The different components are indicated as follows: Northern external lobe (L1), Southern external lobe (L2), Northern internal lobe (L3), Southern internal lobe (L4), and the core (C). The LOFAR 6$''$ beam is shown in the lower-left corner. Right panel: zoom on the core region, showing an overlay of the Pan-STARRS i-band image (in gray scale) and VLASS (in contours). Contours are multiples of the VLASS image RMS, namely RMS$\times$(--3, 3, 4, 5, 6, 7, 8, 9, 10), negative contours are dashed. The VLASS beam is shown in the lower-left corner.}
\label{fig:newRG}
\end{figure*}

During inspection of the sky region in the LoTSS image of GRG J0801.7+4764, we found a new radio source which we name J0802+4711. As evident in the LOFAR image (see figure \ref{fig:newRG}) the source displays two sets of lobes: an internal dimmer one, and an external brighter one (thus resembling the morphology of a double-double radio galaxy). No core is visible at this frequency in between the two set of lobes. The 144 MHz flux density of each individual lobe is reported in table \ref{tab:newRG}. The source extent is roughly 2 arcmin, implying a projected linear size smaller than the GRG threshold of 0.7 Mpc for a redshift up to $z=0.45$. The source is detected in the FIRST and VLASS surveys, where however only the core and external couple of lobes are visible (i.e. the brightest and compact emission in the GHz domain, see figure \ref{fig:newRG}). Finally, the NVSS survey resolve the source into only two components, due to the lower angular resolution (45 arcsec), corresponding to the external+internal lobes regions. Nevertheless, it allow us to estimate a total flux density at 1.4 GHz of 62 mJy. The flux density values extracted for the single components in the LOFAR, FIRST, and VLASS surveys are reported in Table \ref{tab:newRG}. 
In the FIRST image, the core is detected at only $\sim$5 sigma significance (0.76$\pm$0.08 mJy/beam, with an RMS of 0.15 mJy/beam), and thus it is not reported in the FIRST sources catalogue. Nevertheless, its position at the center of a large and complex radio source make the detection more reliable. The location of this central component is RA(J2000)=08:02:43.994 and Dec(J2000)=+47:11:37.325 with a positional uncertainty of around 1.0 arcsec \footnote{To estimate the position error we have used the empirical expression given in \cite{1997ApJ...475..479W}; the value we obtain is in agreement with the values quoted for  point sources at the detection limit of the FIRST catalog}.  
In the LOFAR image at 6 arcsec the core is not distinguishable, its position from FIRST corresponding to a diffuse, extended emission region between the lobes. This results in a plateau of flux density that makes an upper limit estimate not feasible. At higher frequencies (3 GHz), the core is visible in the VLASS survey, with a 10 sigma detection in a position correspondent to the one from FIRST. Its flux density, extracted via Gaussian fitting, is of 2.88$\pm$0.43 mJy. This allows us to obtain a first indication of the spectral index between 1.4 GHz and 3 GHz of $\alpha=1.69$, indicating an inverted spectrum typical of AGN core components. Nevertheless, the different resolution between the two surveys, and the possibility of a variable flux density for the core, make this an approximate estimate only, to be considered with caution. The external couple of lobes is visible as well, although only the most compact regions (hotspots), while the internal lobes are not detected. The fact that the external lobes hotspots are visible at 3 GHz suggests that they have been recently replenished with plasma from the jets, implying that these lobes are the ones linked to the ongoing radio phase. In this case, the internal couple of lobes should be older, or resulting from a lateral expansion of the jet due to a lower density of the medium. Further data are anyway needed to test this hypothesis. 

%%%%%%%%%%%%%%%%%%%%%%%%%%%%%%%%%%%%%%%%%%%%%%%%%%%%%%%%%%%%%%%%%%%%

 \begin{table}
\caption{Radio flux densities at 144 MHz for the different source components of the newly discovered radio galaxy. The last column report the spectral index for the available VLA surveys in the same configuration (FIRST and VLASS), and at similar resolution.}
\begin{tabular}{lccccccc}
\hline
                            &      LOFAR        &  FIRST                &  VLASS            & $\alpha_{1.4}^{3}$    \\
Component                   &      (mJy)        &  (mJy)                &  (mJy)            &                       \\     
\hline
Core                         &  -               &  0.79$\pm$0.08        & 2.88$\pm$0.43     & 1.69         \\
North Lobe (int)             &  54$\pm$11       &  $<$0.45              & $<$0.48           & -            \\
South Lobe (int)             &  15$\pm$3        &  $<$0.45              & $<$0.48           & -            \\
North Lobe (ext)             &  170$\pm$30      &  14.2$\pm$1.4         & 8.20$\pm$1.23     & -0.72        \\
South Lobe (ext)             &  270$\pm$50      &  32.5$\pm$3.2         & 16.2$\pm$2.4      & -0.91        \\
\hline
\end{tabular}
\label{tab:newRG}
\end{table}

%%%%%%%%%%%%%%%%%%%%%%%%%%%%%%%%%%%%%%%%%%%%%%%%%%%%%%%%%%%%%%%%%%%%

Within the FIRST positional error circle, we found only one optical source clearly detected in the SDSS Sky image. The source is stellar-like in the image, and quite dim, with magnitude ranging from 19.4 in $z$ to 21.5 in $u$, and is classified as a star in the SDSS catalogue \citep{2011AJ....141..189V} as well as in the recently release of the TESS catalogue, incorporating also Gaia release 2 data \citep{2019AJ....158..138S}. 
The source is also reported in the unWISE Catalog, containing the positions and fluxes of roughly 2 billion objects observed by the Wide-field Infrared Survey Explorer (WISE) over the full sky \citep{2019ApJS..240...30S}. Fluxes in W1 and W2 bands have been converted into Vega magnitudes following the prescription given in the catalogue: we obtained a W1 and W2 magnitudes of 15.58 and 14.41, respectively, which provide a W1--W2 colour of 1.17, above the line to classify the  source as a possible AGN (see e.g. \citealt{2012ApJ...753...30S}). Finally, we note that an infrared source is also reported in the AllWise catalogue but with a position offset by 3.2 arcsec, and with brighter W1 and W2 magnitudes (difference of 1.2 and 0.5 magnitudes in W1 and W2, respectively); it is possible that this is the same unWISE source discussed above, but better located, and with slightly different fluxes after co-addition of all publicly available 3-5 $\mu$m WISE images (and a more sophisticated data analysis). 

\begin{figure} 
   \includegraphics[width=\columnwidth]{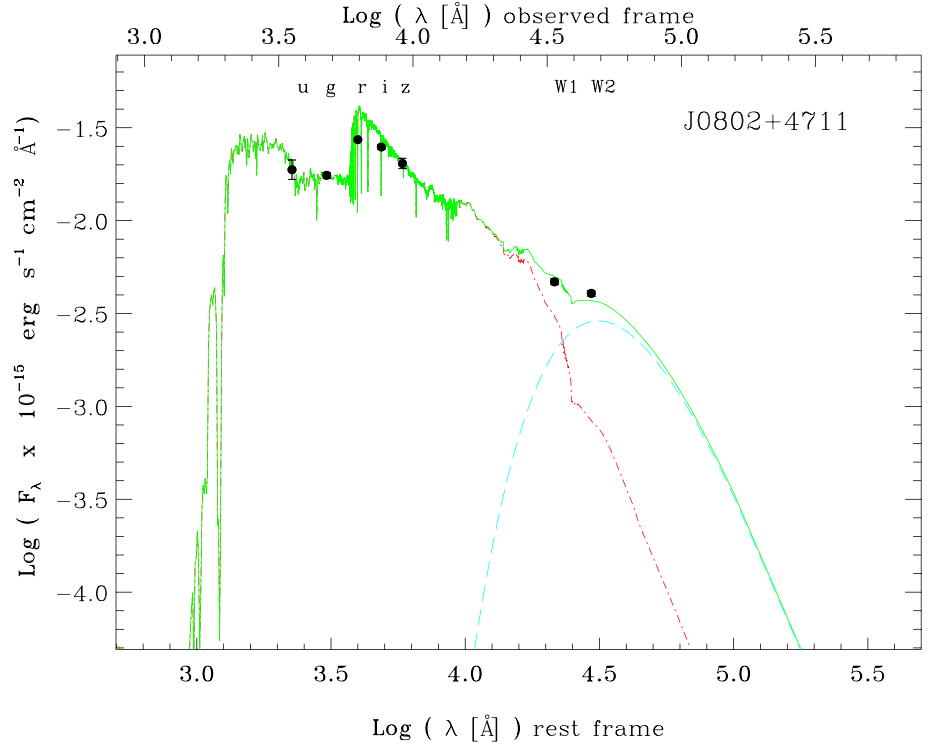} \\
\caption{Optical-infrared (SDSS-WISE) SED of the galaxy J0802+471. The photometric points are fit with the 2SPD fitting code \citep{baldi13,baldi14}: a 0.3-Gyr old stellar population is the red (dot-dashed) line, the dust Black-body component is the green (dashed) lines and the total model is the green solid line. The resulting photometric redshift of the source is 0.57$^{+0.10}_{-0.29}$.  The wavelengths on the top of the plot correspond to observed wavelengths, while those on the bottom are in the rest frame.}
\label{fig:newRG_SED}
\end{figure}

In order to estimate the photometric redshift of the source, we built the optical-infrared SED by using its optical SDSS magnitudes (u, g, r, i, z) and infrared WISE ones (W1 and W2, see figure \ref{fig:newRG_SED}). We used the \verb'2SPD' code, developed by \cite{baldi13} and updated by \cite{baldi14}, to model the optical-infrared SED. This code can include in the SED fit: two stellar populations of different ages, a  dust  component with a single (or two) temperature black-body emission(s),  and an AGN  spectrum  with  emission  lines.  The  code  searches  for the best match between the total (stellar+dust+AGN) models and the photometric points by minimising the appropriate ${\chi}^2$ function (see \citealt{baldi13} for details). The main parameters which 2SPD returns  are  the  photometric  redshift  and  the stellar age  of the host. To estimate the uncertainties  on  the  photo-z  derivation,  we  measure the 99\% confidence solutions. We adopt the synthetic stellar templates from \cite{bruzual11}, with single stellar populations with ages ranging from 1 Myr to 12.5 Gyr. We also use a dust-screen model for the extinction normalised with the free parameter A$_{\rm V}$, and the \cite{2000ApJ...533..682C} extinction law. We run the \verb'2SPD' code for the SED of  J0802+4711 by including a single stellar population and a sigle dust component, which are necessary to reproduce, respectively, the optical and infrared part of the SED. The best fit is consistent with an 0.3 Gyr  old  stellar population (extincted A$_{\rm V}$ = 0.9). The derived photometric  redshift is 0.57$^{+0.10}_{-0.29}$. The total radio power at 1.4 GHz, calculated using the nominal value of the estimated redshift, is $P_{1.4\, GHz}=5.0\times10^{25}$ W/Hz. Even considering the redshift uncertainty, this would place the source in the radio-loud regime according to \cite{1992ARA&A..30..575C}, who defines a threshold of $10^{23}$ W/Hz to distinguish the radio-quiet and radio-loud populations. The angular size as calculated from the LOFAR 20$''$ image is 160 arcsec, that would translate into a projected linear size of 1040 kpc considering the nominal value of the photometric redshift. This would be above the conventional threshold for GRG (700 kpc). However, considering the large redshift uncertainty, the projected linear size could still be as low as 670 kpc. Thus, we cannot consider this object a bona-fide GRG until a more precise redshift estimate will be available. 

Concluding, we found a possible AGN detected at radio, optical, and infrared wavebands, whose nature is still uncertain, and cannot be clarified until the source is either detected in X-rays, or spectroscopically classified. Unfortunately, the source has not been observed at X-ray wavelengths yet, nor it is bright enough for an optical spectroscopic follow-up with small telescopes (more easily available). For this reason, we plan to perform a \emph{Swift}/XRT observation in the near future. %Alternatively, one might assume that the source core is silent both at radio and optical/IR wavelenghts, implying that it has already died out. 
Further radio observations will shed light on this intriguing source.

%%%%%%%%%%%%%%%%%%%%%%%%%%%%%%%%%%%%%%%%%%%%%%%%%%%%%%%%%%%%%%%%%%%%

%%%%%%%%%%%%%%%%%%%%%%%%%%%%%%%%%%%%%%%%%%%%%%%%%%%%%%%%%%%%%%%%%%%%

%%%%%%%%%%%%%%%%%%%%%%%%%%%%%%%%%%%%%%%%%%%%%%%%%%%%%%%%%%%%%%%%%%%
%%%%%%%%%%%%%%%%%%%%%%%%%%%%%%%%%%%%%%%%%%%%%%%%%%%%%%%%%%%%%%%%%%%
% Don't change these lines
\bsp	% typesetting comment
\label{lastpage}
\end{document}